\documentclass[11pt,a4paper]{article}
\pdfoutput=1

\usepackage{jheppub}
\usepackage[utf8]{inputenc}
\usepackage[T1]{fontenc}
\usepackage{afterpage}
\usepackage{enumitem}
\usepackage{bm}

\toccontinuoustrue

\newcommand{\eVq}{\ensuremath{\text{eV}^2}}
\newcommand{\Dmq}{\Delta m^2}
\newcommand{\dd}{\mathrm{d}}
\newcommand{\ringbar}[1]{\mathring{\bar{#1}}}

\DeclareMathOperator{\diag}{\mathbf{diag}}
\DeclareMathOperator{\sinc}{sinc}
\DeclareMathOperator{\rect}{rect}
\DeclareMathOperator{\Tr}{Tr}

\title{From ray to spray: augmenting amplitudes and taming fast
  oscillations in fully numerical neutrino codes}

\author{Michele Maltoni}
\emailAdd{michele.maltoni@csic.es}
\affiliation{Instituto de F\'isica Te\'orica (IFT-CFTMAT), CSIC-UAM,
  Calle de Nicol\'as Cabrera 13--15, Campus de Cantoblanco, E-28049
  Madrid, Spain}

\abstract{In this note we describe how to complement the neutrino
  evolution matrix calculated at a given energy and trajectory with
  additional information which allows to reliably extrapolate it to
  nearby energies or trajectories without repeating the full
  computation.  Our method works for arbitrary matter density
  profiles, can be applied to any propagation model described by an
  Hamiltonian, and exactly guarantees the unitarity of the evolution
  matrix.  As a straightforward application, we show how to enhance
  the calculation of the theoretical predictions for experimentally
  measured quantities, so that they remain accurate even in the
  presence of fast neutrino oscillations.  Furthermore, the ability to
  ``move around'' a given energy and trajectory opens the door to
  precise interpolation of the oscillation amplitudes within a grid of
  tabulated values, with potential benefits for the computation speed
  of Monte-Carlo codes.  We also provide a set of examples to
  illustrate the most prominent features of our approach.}

\preprint{IFT-UAM/CSIC-23-99}


\arxivnumber{2308.00037}

\begin{document}

\maketitle

\section{Motivation}
\label{cha:intro}

The discovery of neutrino oscillations have finally provided robust
observational evidence that the Standard Model of particle physics is
not the ultimate theory of nature.  Lepton flavor conversion requires
neutrinos to be massive, something which was not accounted for in the
original formulation of the Standard Model.  Adding neutrino masses
through the usual Higgs mechanism is of course possible, but involves
the introduction of right-handed neutrino states which, being gauge
singlets, are not prevented by any known symmetry to acquire a
Majorana mass.  Hence, one way or another, New Physics seems to be at
work in the neutrino sector, either through the appearance of
something fundamentally new such as Majorana particles, or through
some unknown mechanisms which prevents them.  It is therefore
understandable that during the last decades an intense experimental
neutrino program has been carried out, and that even more powerful
experiments are being developed for the coming years.

Neutrino experiments, by their very nature, aim at reconstructing
neutrino properties by observing the effects of flavor conversion
during propagation from source to detector.  Such conversion depends
of course on the assumed theoretical model (standard three neutrinos,
extra sterile states, non-standard interactions, \textit{etc.\@}) and
on the specific value of its parameters.  But it also depends on a set
of ``dynamical variables'' characterizing the neutrino state (such as
its energy $E$) or its trajectory (such as the path length $L$, or
more generically the matter profile encountered along the path).  In
what follows we will sometimes refer to a specific instance of these
variables (\textit{i.e.}, a concrete choice of energy, trajectory,
\textit{etc.\@}) as a ``ray''.  Although in principle the experimental
setup aims at determining the dynamical variables as accurately as
possible, so to minimize their impact on the oscillation pattern and
therefore extract the maximum information on the neutrino properties,
in practice some amount of uncertainty is unavoidable.  For example,
the energy spectrum of the neutrinos emitted by the source is usually
non-monochromatic, and the energy resolution of the detector is
finite, so neutrino energy is never perfectly known.  For atmospheric
neutrinos~\cite{Honda:2015fha} the imperfect reconstruction of the
arrival direction implies that the traveled length and crossed matter
profile are uncertain, furthermore the altitude of the production
point is totally unobserved.  Similarly, for solar
neutrinos~\cite{Vinyoles:2016djt} it is impossible to determine at
which point of the solar core the neutrino was produced, and therefore
the exact profile of the traversed matter.  All this implies that the
comparison of experimental results with the theoretical expectations
(be it the $\chi^2$ of the ``number of events'' in a given data bin
defined in terms of reconstructed quantities, or just the likelihood
function for each individual event) always implies integrals (or
averages) over dynamical variables such as the neutrino energy $E$ or
the path length $L$.

Now, these integrals can be performed in many ways.  Let us focus on
the neutrino energy for definiteness, as it is ubiquitous to all
experiments.  One can resort to Riemann integration and divide the
relevant range into a number of small bins, choose a representative
value in each of them, calculate the corresponding conversion
probabilities (for the given theoretical model and parameter values),
and sum.  Or one can use Monte-Carlo techniques, generating a random
set of energy values according to some appropriate distribution.
Either way, the conversion probabilities for each sampled point are
calculated assuming a specific energy value <<$E$>>, but are then used
for the whole interval <<$\dd E$>> which such energy represent.  This
procedure implicitly assumes that the conversion probabilities will
``stay the same'' over the interval <<$\dd E$>>, or at least that
their variation can be reliably inferred from nearby sampled points
without the need of further information.  This is certainly true if
the integration bins are ``small enough'', but sometimes this is
prohibitively difficult to achieve: for example, when oscillations are
``fast'', which requires to choose very small intervals and therefore
a very large number of integration points.

In brief, any point explicitly sampled from the integration domain
(\textit{i.e.}, any ``ray'') actually represent a small but extended
region (a ``spray'') around it.  In view of this, when calculating
neutrino propagation for a given ray one should make sure to recompile
enough information to describe the conversion probabilities in the
whole neighborhood it represents.  In this note we present a method to
address this issue in full generality, \textit{i.e.}, without assuming
a specific oscillation model or matter density profile.  We focus on
scenarios where the evolution of the neutrino state is unitary and can
be described in terms of an hermitian Hamiltonian by means of a
Schrödinger equation.  We do not consider here dissipative processes
such as those commonly accounted by a Lindblad equation, nor neutrino
self-interactions which become relevant in dense media such as
supernovas.  A peculiar feature of our approach is that it is entirely
formulated in terms of generic matrices which play a specific role in
the description of neutrino propagation (such as the Hamiltonian
$\bm{H}$, the evolution matrix $\bm{S} \equiv \mathcal{T}\! \exp[-i
  \int\! \bm{H}\, \dd x]$, \textit{etc.\@}) but whose concrete
expression as a function of the parameters of the model is never taken
into account.  This happens because we do not aim at providing
analytic formulas valid for specific oscillation scenarios (which is
already a widely studied topic in the literature, see
refs.~\cite{Arafune:1996bt, Lisi:1997yc, Arafune:1997hd,
  Minakata:1998bf, Petcov:1998su, Akhmedov:1998ui, Akhmedov:1998xq,
  Chizhov:1999he, Ohlsson:1999xb, Koike:1999tb, Ohlsson:1999um,
  Harrison:1999df, DeRujula:2000ap, Cervera:2000kp, Lunardini:2000swa,
  Minakata:2000ee, Akhmedov:2000cs, Lisi:2000su, Ohlsson:2001et,
  Freund:2001pn, Akhmedov:2001kd, Yasuda:2001va, Kimura:2002hb,
  Kimura:2002wd, Yokomakura:2002av, Aguilar-Arevalo:2003hty,
  Jacobson:2003wc, Akhmedov:2004ny, deHolanda:2004fd,
  Ioannisian:2004jk, Akhmedov:2004rq, Blennow:2004qd,
  Ioannisian:2004vv, Akhmedov:2005yj, Takamura:2005df, Choubey:2005zy,
  Akhmedov:2006hb, deAquino:2007sx, Liao:2007re, Ioannisian:2008ve,
  Akhmedov:2008qt, Kikuchi:2008vq, Asano:2011nj, Agarwalla:2013tza,
  Blennow:2013rca, Coloma:2014kca, Xu:2015kma, Minakata:2015gra,
  Parke:2016joa, Denton:2016wmg, Ge:2016dlx, Fong:2016yyh, Li:2016pzm,
  Pas:2016qbg, Minakata:2017ahk, Fong:2017gke, Denton:2018hal,
  Ioannisian:2018qwl, Denton:2018fex, Bernabeu:2018twl,
  Martinez-Soler:2018lcy, Bernabeu:2018use, Li:2018ezt,
  Denton:2018cpu, Chaves:2018sih, Barenboim:2019pfp, Denton:2019yiw,
  Martinez-Soler:2019nhb, Parke:2019jyu, Denton:2019ovn,
  Martinez-Soler:2019noy, Wang:2019dal, Denton:2019qzn, Luo:2019efb,
  Huber:2019frh, Ioannisian:2020isl, Minakata:2020ijz,
  Minakata:2020oxb, Parke:2020wha, Minakata:2021dqh, Denton:2021vtf,
  Minakata:2021goi, Minakata:2022yvs, Abdullahi:2022fkh,
  Denton:2023zwa, Minakata:2023ice} for an incomplete list), but
rather at developing a model-independent framework which could serve
as a guideline to enrich existing algorithms and extend their range of
applicability ``from the ray to the spray''.

This work is organized as follows.  In chapter~\ref{cha:taylor} we
describe how to complement the conversion amplitudes calculated at a
given energy and trajectory with extra information which allows to
extrapolate them accurately to a neighborhood of such ray.  In
chapter~\ref{cha:average} we apply this formalism to the case of
``fast'' neutrino oscillations, showing how the corresponding
averaging effects can be implemented in fully numerical calculations.
In chapter~\ref{cha:examples} we provide a set of examples to
illustrate the advantages and limitations of the proposed approach,
and in chapter~\ref{cha:summary} we summarize our conclusions.
Finally, in appendix~\ref{app:splitting} we briefly discuss how
specific symmetries of the neutrino system can be efficiently
exploited within our formalism.

\section{Formalism}
\label{cha:taylor}

The simplest and best known scenario accounting for leptonic flavor
conversion consists in mass-induced neutrino oscillations in vacuum.
From the phenomenological point of view, the fundamental properties of
such model are:
\begin{enumerate}[label=\emph{\alph*})]
\item\label{it:vacuum-E} the evolution Hamiltonian $\bm{H}_0$ is
  inversely proportional to the neutrino energy $E$.  This in
  particular implies that $[\bm{H}_0(E_1), \bm{H}_0(E_2)] = 0$ even
  for $E_1 \ne E_2$, so that it exists a basis in flavor space (the
  mass basis) where $\bm{H}_0$ is diagonal for \emph{all} energies;

\item\label{it:vacuum-L} due to translational invariance of vacuum,
  $\bm{H}_0$ is independent of the neutrino position in space, so that
  the evolution matrix $\bm{S}_0 \equiv \exp(-i \bm{H}_0 L)$ depends
  on the neutrino trajectory only through its total length $L$.
\end{enumerate}
In this case the oscillation probabilities take a very simple form,
essentially reducing to the sum of terms proportional to
$\cos(\gamma_i L/E)$ or $\sin(\gamma_i L/E)$ where $\gamma_i$
generically denotes appropriate functions of the oscillation
parameters.  This simplicity allows for an analytic treatment of
various oscillation effects which would otherwise be hard to handle in
a fully numerical framework.  For example, the \textsf{GLoBES}
software~\cite{Huber:2004ka, Huber:2007ji} implements a feature called
``low-pass filter'' which averages probabilities and suppresses
aliasing in the presence of very fast neutrino oscillations, for
neutrino trajectories exhibiting translational invariance
(\textit{i.e.}, vacuum and constant density).
A similar functionality is also provided by the \textsf{nuSQuIDS}
toolbox~\cite{Arguelles:2021twb}, again assuming a spatially uniform
environment.
In general, the two properties~\ref{it:vacuum-E} and~\ref{it:vacuum-L}
above are responsible for the particularly simple dependence of
$\bm{S}_0$ on the neutrino energy and position, respectively.
In the rest of this chapter we will show how, by means of suitable
first-order Taylor expansions of the generic Hamiltonian $\bm{H}$ and
the evolution matrix $\bm{S}$, it is possible to attain such
simplicity also in the general case of fully numerical neutrino
propagation in an arbitrary matter profile, so that some analytic
techniques commonly used for vacuum oscillations become applicable.

\subsection{The constant-density case}
\label{sec:onelayer}

Let's start by considering neutrinos propagating over a distance $L$
in a constant matter potential, so that the evolution Hamiltonian
$\bm{H}(E)$ does not depend on the position.  It is reasonable to
assume that $\bm{H}$ is a smooth function of $E$, so we can expand it
at first order in a neighborhood of a reference energy $\bar{E}$:
\begin{equation}
  \bm{H}(\bar{E} + \xi_E)
  \approx \bar{\bm H} + \bm{H}_E'\, \xi_E
  \quad\text{with}\quad
  \bar{\bm H} \equiv \bm{H}(\bar{E})
  \quad\text{and}\quad
  \bm{H}_E' \equiv
  \left.\frac{\dd\bm{H}(\bar{E} + \xi_E)}{\dd\xi_E} \right|_{\xi_E=0} .
\end{equation}
With this, the matrix $\bm{S}(E)$ describing the neutrino propagation
is:
\begin{equation}
  \bm{S}(E) \equiv e^{-i \bm{H}(E) L}
  \quad\Rightarrow\quad
  \bm{S}(\bar{E} + \xi_E)
  \approx e^{-i (\bar{\bm H} + \bm{H}_E'\, \xi_E) L} \,.
\end{equation}
Now, it would be convenient to factorize $\bm{S}(\bar{E} + \xi_E)$
into the product of two terms, one describing the evolution at the
reference energy, $\bar{\bm S} \equiv \bm{S}(\bar{E})$, and the other
accounting for the perturbation induced by the energy shift $\xi_E$.
In other words, we are seeking an expression of the kind:
\begin{equation}
  \label{eq:factorS}
  \bm{S}(\bar{E} + \xi_E)
  \approx \bar{\bm S}\, e^{-i \bm{K}_E\, \xi_E}
  \quad\text{with}\quad
  \bar{\bm S} = e^{-i \bar{\bm H} L}
\end{equation}
for a suitable matrix $\bm{K}_E$, which can formally be defined as:
\begin{equation}
  \label{eq:defineK}
  \bm{K}_E = i\, \bar{\bm S}^\dagger \cdot
  \left.\frac{\dd\bm{S}(\bar{E} + \xi_E)}{\dd\xi_E} \right|_{\xi_E=0} .
\end{equation}
If the matrices $\bar{\bm H}$ and $\bm{H}_E'$ commute, as is the case
in vacuum, it is immediate to see that $\bm{K}_E$ is proportional to
the derivative of the Hamiltonian, $\bm{K}_E = \bm{H}_E' L$.  In the
general case the expression for $\bm{K}_E$
reads~\cite{Delledalle:2022xx}:
\begin{equation}
  \label{eq:buildK}
  \bar{\bm U}^\dagger\, \bm{K}_E\, \bar{\bm U} \equiv L\,
  \big( \bar{\bm U}^\dagger\, \bm{H}_E'\, \bar{\bm U} \big)
  \odot \bm{C}
  \quad\text{with}\quad
  \bm{C}_{ij} \equiv
  \frac{\exp\big[i(\bar\omega_i - \bar\omega_j)L\big] - 1}
       {i(\bar\omega_i - \bar\omega_j)L}
\end{equation}
where we have introduced the unitary matrix $\bar{\bm U}$ relating the
flavor basis to the effective mass basis at reference energy
$\bar{E}$:
\begin{equation}
  \label{eq:massbasis}
  \bar{\bm U}^\dagger\, \bar{\bm H}\, \bar{\bm U}
  = \bar{\bm\omega}
  \quad\text{with}\quad
  \bar{\bm\omega} = \diag\{ \bar\omega_i \}
  \quad\Rightarrow\quad
  \bar{\bm S}
  = \bar{\bm U} \, e^{-i \bar{\bm\omega} L} \, \bar{\bm U}^\dagger \,.
\end{equation}
In eq.~\eqref{eq:buildK} the operator $\odot$ denotes the Hadamard
product of two matrices, which consists in an element-wise
multiplication of the corresponding elements: $[\bm{A} \odot
  \bm{B}]_{ij} \equiv \bm{A}_{ij}\, \bm{B}_{ij}$.  The matrix $\bm{C}$
is hermitian and its diagonal entries are equal to $1$, while the
non-diagonal entries are no larger than $1$ in modulus: $\bm{C}_{ii} =
1$ and $|\bm{C}_{ij}| \le 1$.  The matrix $\bm{K}_E$ is also
hermitian, which ensures that $\bm{S}(\bar{E} + \xi_E)$ is exactly
unitary for any value of $\xi_E$.  If $[\bar{\bm H}, \bm{H}_E'] = 0$
then it is possible to choose $\bar{\bm U}$ so that $\bar{\bm H}$ and
$\bm{H}_E'$ are simultaneously diagonalized, in which case the
element-wise multiplication by $\bm{C}$ has no effect and $\bm{K}_E =
\bm{H}_E' L$ as previously stated.

From the computational point of view, the most time-consuming step in
the determination of $\bm{K}_E$ is the diagonalization of $\bar{\bm
  H}$~\cite{Hahn:2006hr, Kopp:2006wp}.  However, such a step renders
the computation of $\bar{\bm S}$ (which requires to perform a matrix
exponential) essentially trivial, so that the time spent in
diagonalizing $\bar{\bm H}$ is recovered from its exponentiation.  We
find therefore that computing also $\bm{K}_E$ does not result in
significant slow-down with respect to computing only $\bar{\bm S}$.

\subsection{Multiple layers and arbitrary matter profile}
\label{sec:multilay}

In the previous section we have seen that, in the case of constant
matter density, neutrino oscillations around a central energy
$\bar{E}$ can be described in terms of two matrices: a unitary one
$\bar{\bm S}$, describing the evolution for the specific value $E =
\bar{E}$, and an hermitian one $\bm{K}_E$, allowing to extrapolate
$\bar{\bm S}$ to nearby energies $E = \bar{E} + \xi_E$.  Let us now
consider the case in which the neutrino crosses \emph{two} consecutive
layers, each one with its own matter density.  We will denote by
$(\bar{\bm S}_1, \bm{K}_1^E)$ the evolution and perturbation matrices
of the first layer, and by $(\bar{\bm S}_2, \bm{K}_2^E)$ those of the
second layer.  The combined evolution reads:
\begin{equation}
  \bm{S}(\bar{E} + \xi_E) \approx
  \big[ \bar{\bm S}_2\, e^{-i \bm{K}_2^E\, \xi_E} \big] \cdot
  \big[ \bar{\bm S}_1\, e^{-i \bm{K}_1^E\, \xi_E} \big] \,.
\end{equation}
The expression for $\bar{\bm S} \equiv \bm{S}(\bar{E})$ is readily
obtained by setting $\xi_E = 0$ in the previous formula, which yields
$\bar{\bm S} = \bar{\bm S}_2 \cdot \bar{\bm S}_1$ as expected.  As for
the combined $\bm{K}_E$, it can be found by means of
eq.~\eqref{eq:defineK}, which gives:
\begin{align}
  \bm{K}_E
  &= i \big[ \bar{\bm S}_2\, \bar{\bm S}_1 \big]^\dagger
  \cdot \bigg[
    \bar{\bm S}_2\, e^{-i \bm{K}_2^E \xi_E} (-i \bm{K}_2^E)
    \bar{\bm S}_1\, e^{-i \bm{K}_1^E \xi_E}
    +
    \bar{\bm S}_2\, e^{-i \bm{K}_2^E \xi_E}
    \bar{\bm S}_1\, e^{-i \bm{K}_1^E \xi_E} (-i \bm{K}_1^E)
    \bigg]_{\xi_E=0}
  \nonumber
  \\
  &= \big[ \bar{\bm S}_1^\dagger \, \bar{\bm S}_2^\dagger \big]
  \cdot \big[
    \bar{\bm S}_2\, \bm{K}_2^E\, \bar{\bm S}_1 +
    \bar{\bm S}_2 \, \bar{\bm S}_1\, \bm{K}_1^E
    \big]
  = \bar{\bm S}_1^\dagger \, \bm{K}_2^E\, \bar{\bm S}_1
  + \bm{K}_1^E \,.
\end{align}
These two expressions can be summarized in a single ``multiplication
rule'' among the \emph{pairs} of $(\bar{\bm S}, \bm{K}_E)$ matrices
characterizing each layer:
\begin{equation}
  \label{eq:pairmult}
  (\bar{\bm S}_2,\, \bm{K}_2^E) \cdot (\bar{\bm S}_1,\, \bm{K}_1^E)
  \to (\bar{\bm S}_2 \, \bar{\bm S}_1,~
  \bar{\bm S}_1^\dagger \, \bm{K}_2^E\, \bar{\bm S}_1 + \bm{K}_1^E) \,.
\end{equation}
Interestingly, this product is associative, it has $(\bm{1}, \bm{0})$
as identity element, and every pair $(\bar{\bm S}, \bm{K}_E)$ has
$(\bar{\bm S}^\dagger, -\bar{\bm S} \bm{K}_E \bar{\bm S}^\dagger)$ as
inverse, so that the set of pairs form a group.  This is not a
surprise, since it is clear from eq.~\eqref{eq:factorS} that the
introduction of $\bm{K}_E$ does not alter the unitarity (and therefore
the algebraic structure) of the evolution matrix $\bm{S}(E)$.

From eq.~\eqref{eq:pairmult} it is evident that the extension to
trajectories with varying matter potential of the formalism developed
in the previous section follows the same line as the usual
``fixed-energy'' calculation of the evolution matrix $\bar{\bm S}$,
except that this one is now replaced by a $(\bar{\bm S}, \bm{K}_E)$
pair.  Concretely, we proceed as follows:
\begin{itemize}
\item we divide the trajectory into $N$ smaller layers, in such a way
  that the variation of the matter potential within each of them is
  small compared to its average value;

\item we calculate the evolution pair $(\bar{\bm S}_n, \bm{K}_n^E)$ for
  all the $n = 1, \dots, N$ layers, under the hypothesis of constant
  matter density;

\item we merge together all the individual pairs using the product
  defined in eq.~\eqref{eq:pairmult}, so that the overall evolution
  pair is $(\bar{\bm S}, \bm{K}_E) = (\bar{\bm S}_N, \bm{K}_N^E) \cdot
  ({}\cdots) \cdot (\bar{\bm S}_1, \bm{K}_1^E)$.
\end{itemize}
As for the constant-density case, no significant amount of extra time
is required to compute also $\bm{K}_E$ as compared to computing only
$\bar{\bm S}$, provided that the time for matrix multiplication is
negligible with respect to that of matrix exponentiation.

\subsection{Perturbations of the neutrino trajectory}
\label{sec:zenith}

So far we have only considered perturbations of the evolution matrix
$\bar{\bm S}$ around a reference energy $\bar{E}$.  However, sometimes
the calculation of the event rates may require additional integration
over other dynamical variables: for example, for atmospheric neutrinos
the arrival direction (parametrized by the zenith angle $\Theta$)
plays a key role.  In these cases, the same first-order expansion
which we have just presented for the neutrino energy can be repeated
for the other integration variables $X$, by considering the
corresponding $\bm{K}_X$ matrices.  For instance, in the case of
atmospheric neutrinos, for each reference energy $\bar{E}$ and zenith
angle $\bar{\Theta}$ we will write:
\begin{equation}
  \label{eq:factorSatm}
  \bm{S}(\bar{E} + \xi_E, \bar\Theta + \xi_\Theta)
  \approx \bar{\bm S}\, e^{-i \bm{K}_E\, \xi_E}\, e^{-i \bm{K}_\Theta\, \xi_\Theta}
  \quad\text{with}\quad
  \bar{\bm S} \equiv \bm{S}(\bar{E}, \bar\Theta) \,.
\end{equation}
Notice that, although strictly speaking the final result depends on
the \emph{ordering} in which we introduce the perturbation factors
$e^{-i \bm{K}_E\, \xi_E}$ and $e^{-i \bm{K}_\Theta\, \xi_\Theta}$ (as
in the general case the matrices $\bm{K}_E$ and $\bm{K}_\Theta$ may
not commute), the effect of interchanging them is of order $\xi_E\,
\xi_\Theta$, and can therefore be neglected in our first-order
expansion.

For what concerns neutrinos crossing the Earth, the construction of
$\bm{K}_\Theta$ is particularly simple.  Given the spherical symmetry
of the Earth, it is possible to approximate its density profile with a
large number of constant-density shells.  A given trajectory will
cross a specific sequence $n = 1, \dots, N$ of such shells, each one
with a length $L_n(\Theta)$ determined by the geometry.  A
\emph{large} variation of $\Theta$ will cause shells to drop in or out
of the reference path $\bar\Theta$ (\textit{i.e.}, a change in $N$),
and this is a non-analytic effect which no Taylor expansion can
reproduce.  But for trajectories close enough to the central one, the
sequence of shells (and therefore of the Hamiltonians $\bar{\bm H}_n$
used within each of them) will not change, only the traveled length in
each shell will be affected.  Therefore:
\begin{equation}
  \bm{K}_n^\Theta = \bar{\bm H}_n \left.
  \frac{\dd L_n(\bar\Theta + \xi_\Theta)}{\dd\xi_\Theta}
  \right|_{\xi_\Theta=0}
\end{equation}
where we have taken advantage of the fact that the ``perturbation''
commutes with the Hamiltonian $\bar{\bm H}_n$, so that no Hadamard
product with a matrix $\bm{C}$ is required in this case.  The
$(\bar{\bm S}_n, \bm{K}_n^E, \bm{K}_n^\Theta)$ matrices for the
various layers can then be merged together using the composition rule
in eq.~\eqref{eq:pairmult}, trivially extended to accommodate also the
$\Theta$ derivative:
\begin{equation}
  \label{eq:tuplemult}
  (\bar{\bm S}_2,\, \bm{K}_2^E,\, \bm{K}_2^\Theta) \cdot
  (\bar{\bm S}_1,\, \bm{K}_1^E,\, \bm{K}_1^\Theta)
  \to (\bar{\bm S}_2 \, \bar{\bm S}_1, \enspace
  \bar{\bm S}_1^\dagger \, \bm{K}_2^E\, \bar{\bm S}_1 + \bm{K}_1^E, \enspace
  \bar{\bm S}_1^\dagger \, \bm{K}_2^\Theta\, \bar{\bm S}_1 + \bm{K}_1^\Theta) \,.
\end{equation}
It is clear that further integration variables $X$ which may affect
neutrino propagation can be handled in the same way by simply
appending their own perturbation matrices $\bm{K}_X$ to the tuples of
eq.~\eqref{eq:tuplemult}.  A particularly simple situation occur when
$X$ accounts for a longitudinal extension of the neutrino trajectory
at one of its extremes, as implied by averaging over an extended
production region or a non-negligible detector volume.  In this case
the matrix $\bm{K}_X$ for the \emph{entire} trajectory is directly
related to the concrete value assumed by the evolution Hamiltonian at
the relevant extreme, namely $\bm{K}_X = \bm{H}_\text{src}$ at the
source or $\bm{K}_X = \bar{\bm S}^\dagger\, \bm{H}_\text{det}\,
\bar{\bm S}$ at the detector.

\subsection{Improved evolution within a definite layer}
\label{sec:improved}

Till now we have discussed how to account for small deviations of
dynamical variables (such as neutrino energy $E$ or zenith angle
$\Theta$) from a central value used to perform the actual
calculations.  In this section we will instead concentrate on the
reference ray itself (defined as $E = \bar{E}$ and $\Theta =
\bar\Theta$) and in particular on the construction of its evolution
matrix (denoted as $\bar{\bm S}$ in previous sections), showing how
first-order Taylor expansion can be used to improve its computation as
well.

As seen in section~\ref{sec:multilay}, a generic way to handle
neutrino propagation in an arbitrary matter profile it to divide the
trajectory into a number of layers, small enough so that the variation
of the matter density within each of them can be considered small.  We
will focus here on one of these layers, assumed to have length $L$,
and parametrize by $x \in [0,L]$ the instantaneous neutrino position
inside it.  Let us denote by $\bm{S}(x)$ the unitary matrix describing
the transition and survival amplitudes of the neutrino state from the
beginning of the layer to position $x$, so that $\bm{S}(0) = \bm{1}$
whereas $\bm{S}(L)$ corresponds to the evolution over the entire
layer.  The matrix $\bm{S}(x)$ satisfies the same equation as the
state vector:
\begin{equation}
  \label{eq:smatreq}
  i \frac{\dd\bm{S}(x)}{\dd x} = \bm{H}(x) \, \bm{S}(x) \,.
\end{equation}
By construction, the matter density varies little within the layer.  A
zero-order approximation is therefore to assume that it is perfectly
constant, as we did in section~\ref{sec:onelayer}.  In this case
$\bm{H}(x) = \bar{\bm H}$ and eq.~\eqref{eq:smatreq} can be solved
immediately:
\begin{equation}
  i \frac{\dd\bar{\bm S}(x)}{\dd x} = \bar{\bm H} \, \bar{\bm S}(x)
  \quad\Rightarrow\quad
  \bar{\bm S}(x) = e^{-i \bar{\bm H} x}
  = \bar{\bm U} \, e^{-i \bar{\bm\omega} x} \, \bar{\bm U}^\dagger
\end{equation}
where $\bar{\bm\omega} = \diag\{ \bar\omega_i \} = \bar{\bm
  U}^\dagger\, \bar{\bm H}\, \bar{\bm U}$ as defined in
eq.~\eqref{eq:massbasis}.  For convenience we have denoted by
$\bar{\bm S}(x)$ the solution in the constant density approximation.
The evolution matrix $\bar{\bm S}(L)$ of the entire layer is:
\begin{equation}
  \label{eq:Slayer0}
  \bar{\bm S}(L)
  = \bar{\bm U} \, e^{-2i \bar{\bm\phi}} \, \bar{\bm U}^\dagger
  \quad\text{with}\quad
  \bar{\bm\phi} \equiv \diag\{ \bar\phi_i \}
  \quad\text{and}\quad
  \bar\phi_i \equiv \frac{\bar\omega_i L}{2}
\end{equation}
in full agreement with the formalism of section~\ref{sec:onelayer},
and in particular with the definition of $\bar{\bm S}$ appearing in
eqs.~\eqref{eq:factorS} and~\eqref{eq:massbasis}.

The purpose of this section is to go beyond the constant-density
approximation.  To this aim, let us now introduce a small perturbation
$\bm\Delta(x)$ and decompose the Hamiltonian $\bm{H}(x)$ within our
layer as:
\begin{equation}
  \label{eq:decomp}
  \bm{H}(x) = \bar{\bm H} + \bm\Delta(x)
\end{equation}
where $\bar{\bm H}$ can be chosen as the value of $\bm{H}(x)$ at some
specific location (such as $x=0$ or $x=L/2$) or be defined by the
condition $\langle \bm\Delta \rangle \equiv \int_0^L \bm\Delta(x) \,
dx \mathbin{\big/} L = \bm{0}$.  Following the approach of
ref.~\cite{Akhmedov:2005yj}, we seek the solution of
eq.~\eqref{eq:smatreq} in the form
\begin{equation}
  \label{eq:exp2}
  \bm{S}(x) = \bar{\bm S}(x) \, e^{-i\bm{K}(x)} \approx
  \bar{\bm S}(x) \left[ \bm{1} - i \bm{K}(x) \right] \,,
\end{equation}
with $\bm{K}(x)$ satisfying $|K_{ab}(x)| \ll 1$.  Inserting
eq.~\eqref{eq:exp2} into eq.~\eqref{eq:smatreq} and keeping only the
first order terms in $\bm\Delta(x)$ and $\bm{K}(x)$, we find:
\begin{equation}
  \label{eq:diff}
  \frac{\dd\bm{K}(x)}{\dd x} =
  \bar{\bm S}^\dagger(x) \, \bm\Delta(x) \, \bar{\bm S}(x) \,.
\end{equation}
At this point it is convenient to switch to the effective mass basis,
so that $\bar{\bm H}$ and $\bar{\bm S}(x)$ become diagonal.  Defining:
\begin{equation}
  \tilde{\bm K}(x)
  \equiv \bar{\bm U}^\dagger \, \bm{K}(x) \, \bar{\bm U} \,,
  \quad
  \tilde{\bm\Delta}(x)
  \equiv \bar{\bm U}^\dagger \, \bm\Delta(x) \, \bar{\bm U}
\quad\Rightarrow\quad
  \frac{\dd\tilde{\bm K}(x)}{\dd x} =
  e^{i \bar{\bm\omega} x} \, \tilde{\bm\Delta}(x) \, e^{-i \bar{\bm\omega} x}
\end{equation}
we see that the differential equation for $\tilde{\bm K}(x)$ separates
into individual components, and can therefore be solved by ordinary
integration:
\begin{equation}
  \label{eq:KtildeG}
  \tilde{\bm K}_{ij}(L) = \int_0^L e^{i (\bar\omega_i - \bar\omega_j) x} \,
  \tilde{\bm\Delta}_{ij}(x) \, \dd x \,.
\end{equation}
This expression is just a special case of the general formalism
presented in ref.~\cite{Ioannisian:2008ve}, and corresponds to the
truncation of the Magnus series to its first term.  Specific
derivations for concrete matter density profiles can be found in the
literature, for example the Earth structure predicted by the PREM
model~\cite{Dziewonski:1981xy} involves density shells which are well
described by eq.~\eqref{eq:decomp}.  Accounting for the perturbation
$\bm\Delta(x)$ on top of the constant part $\bar{\bm H}$ allows to
compute the neutrino evolution inside each Earth shell in a single
shot, without the need to further subdivide it into smaller layers.
Examples of this approach can be found, \textit{e.g.}, in
refs.~\cite{Lisi:1997yc, Gonzalo:2023mdh} for solar neutrinos and in
ref.~\cite{Akhmedov:2006hb} for atmospheric neutrinos.

In the present work, however, we do not want to stick to concrete
matter density profiles, but we are interested instead in formulas
which can be applied to generic situations.  If the constant-density
limit can be regarded as a zero-order approximation, then the natural
first-order generalization is to assume that $\bm\Delta(x)$ is a
linear function of $x$ within the given layer~\cite{deAquino:2007sx}:
$\bm\Delta(x) = (x - L/2)\, \bm{H}'$.  In such case the integral in
eq.~\eqref{eq:KtildeG} can be computed analytically and we get:
\begin{equation}
  \label{eq:KtildeL}
  \tilde{\bm K}_{ij}(L)
  = L^2 \, e^{i(\bar\phi_i - \bar\phi_j)} \,
  \tilde{\bm H}_{ij}' \,
  \hat{\bm C}_{ij}
  \quad\text{with}\quad
  \hat{\bm C}_{ij} \equiv \frac{\sinc'(\bar\phi_i - \bar\phi_j)}{2i}
\end{equation}
where $\sinc'(x)$ denotes the first derivative of the unnormalized
$\sinc(x)$ function:
\begin{equation}
  \label{eq:sincs}
  \sinc(x) \equiv \frac{\sin(x)}{x} \,,
  \qquad
  \sinc'(x) \equiv \frac{\dd\sinc(x)}{\dd x}
  = \frac{\cos(x) - \sinc(x)}{x} \,.
\end{equation}
Switching back to the flavor basis and using matrix notation,
eq.~\eqref{eq:KtildeL} becomes:
\begin{equation}
  \label{eq:Khat}
  \bm{K}(L)
  = \bar{\bm U} \, e^{i\bar{\bm\phi}} \, \hat{\bm K} \,
  e^{-i\bar{\bm\phi}} \, \bar{\bm U}^\dagger
  \quad\text{with}\quad
  \hat{\bm K} \equiv L^2\, \big( \bar{\bm U}^\dagger\, \bm{H}'\,
  \bar{\bm U} \big) \odot \hat{\bm C} \,.
\end{equation}
In principle the factor $e^{i(\bar\phi_i - \bar\phi_j)}$ in
eq.~\eqref{eq:KtildeL} could have been included into the definition of
$\hat{\bm C}_{ij}$, in which case the phase matrices $e^{\pm
  i\bar{\bm\phi}}$ in eq.~\eqref{eq:Khat} would not have appeared.  In
either case the matrix $\hat{\bm C}$ is hermitian and with zero
diagonal entries; but with the present choice $\hat{\bm C}$ is also
purely imaginary, which helps to speed up calculations when both
$\bar{\bm H}$ (and therefore $\bar{\bm U}$) and $\bm{H}'$ are real
matrices.  In our convention the expression for the evolution matrix
$\bm{S}(L)$ of the entire layer, including the first-order correction
$\hat{\bm K}$, reads:
\begin{equation}
  \label{eq:Slayer1}
  \bm{S}(L)
  = \bar{\bm S}(L) \, e^{-i \bm{K}(L)}
  =  \bar{\bm U} \, e^{-i \bar{\bm\phi}} \, e^{-i \hat{\bm K}} \,
  e^{-i \bar{\bm\phi}} \, \bar{\bm U}^\dagger \,.
\end{equation}
Notice that the matrix $\hat{\bm K}$ defined in eq.~\eqref{eq:Khat}
plays a different role than the perturbation matrices $\bm{K}_E$ and
$\bm{K}_\Theta$ introduced in the previous sections.  As detailed in
eq.~\eqref{eq:factorSatm}, $(\bm{K}_E, \bm{K}_\Theta)$ describe how to
alter the evolution matrix $\bar{\bm S}$ when the dynamical variables
$(E, \Theta)$ deviate from their reference values $(\bar{E},
\bar\Theta)$ by finite amounts $(\xi_E, \xi_\Theta)$.  In turn,
$\hat{\bm K}$ encodes a correction to the constant-density
approximation which is not controlled by any tunable quantity, and
therefore there is no reason to keep it separated from $\bar{\bm S}$.
For this reason, the correct way to implement $\hat{\bm K}$ into the
formalism developed in the previous sections is simply to replace
$\bar{\bm S} \equiv \bar{\bm S}(L)$ from eq.~\eqref{eq:Slayer0} with
$\bar{\bm S} \equiv \bm{S}(L)$ from eq.~\eqref{eq:Slayer1} in the
construction of the tuple $(\bar{\bm S}, \bm{K}_E, \bm{K}_\Theta)$
characterizing neutrino propagation.  As for trajectories with
multiple layers, one simply repeats this procedure for each layer and
then combines them together using the multiplication rule of
eq.~\eqref{eq:tuplemult}.

\section{Averaging}
\label{cha:average}

In chapter~\ref{cha:taylor} we have presented a formalism which allows
to calculate the neutrino transition amplitudes in an extended
neighborhood of a reference energy and trajectory.  Here we will make
use of these results to derive expressions for the flavor conversion
probabilities, which are the key ingredient in the calculation of the
theoretical predictions for experimentally measured quantities.  In
particular, we will show how our approach ensures that the integrals
over the dynamical variables (such as the neutrino energy or
trajectory) remain accurate even in the presence of fast neutrino
oscillations, avoiding aliasing without the need to increase the
density of integration points.

The number of events observed by a neutrino experiment can be usually
written as the sum over many oscillation channels (corresponding to
initial flavor, final flavor, neutrino chirality, and so on) of
expressions of the form:
\begin{equation}
  \label{eq:integral}
  N_\text{ch} \propto \int N(E)\, P(E)\, \dd E
\end{equation}
where $P(E)$ is the neutrino conversion probability for the given
oscillation channel, and $N(E)$ denotes the ``unoscillated number of
events'' which takes into account the neutrino flux at the source, the
cross-section of the process, the efficiency and finite resolution of
the detector, the number of targets and running time, and in general
every factor or function which is required to properly describe the
experimental setup.  In principle, the integral in
eq.~\eqref{eq:integral} should extend over all the dynamical variables
which affect neutrino propagation, such as the arrival direction or
the production point for extended sources, but for definiteness we
focus here only on the neutrino energy $E$.

In order to evaluate the integral numerically, it is useful to divide
the integration domain into small intervals $[E_i,\, E_{i+1}]$, so
that $N_\text{ch} = \sum_i N_\text{ch}^i$ with:
\begin{equation}
  \label{eq:binned}
  N_\text{ch}^i \propto \int_{E_i}^{E_{i+1}} N(E)\, P(E)\, \dd E
  = \Delta_E \cdot \langle N P \rangle
  \quad\text{with}\quad
  \Delta_E \equiv E_{i+1} - E_i
\end{equation}
where $\langle\,\rangle$ denotes the average over the given bin,
\textit{i.e.}, the integral itself divided by the bin's width
$\Delta_E$.  In what follows we will assume that the function $N(E)$
is relatively ``smooth'', in the sense that within each energy
interval it is well approximated by a straight line:
\begin{equation}
  N(\bar{E} + \xi_E) \approx
  \bar{N} + N_E'\, \xi_E
  \quad\text{with}\quad
  \bar{E} \equiv (E_i + E_{i+1} ) \mathbin{\big/} 2
  \quad\text{and}\quad
  \xi_E \in \bigg[-\frac{\Delta_E}{2},\, +\frac{\Delta_E}{2}\bigg] .
\end{equation}
In turn, while we assume that the probability $P(E)$ is continuous and
differentiable, we do not require that it exhibits such slow
variation, at least not so for every point in the parameter space.
With this, eq.~\eqref{eq:binned} becomes:
\begin{equation}
  \label{eq:taylor}
  N_\text{ch}^i \propto \bigg[
  \Delta_E \cdot \bar{N} \cdot \langle P \rangle
  + \Delta_E^2 \cdot N_E' \cdot
  \bigg\langle P \cdot \frac{\xi_E}{\Delta_E} \bigg\rangle
  + \mathcal{O}(\Delta_E^3) \bigg] \,.
\end{equation}
Since $0 \le P(E) \le 1$ and $|\xi_E / \Delta_E| \le 1/2$, the average
$\langle P \cdot \xi_E / \Delta_E \rangle$ is no larger than $1/2$ in
absolute value, so the second term in eq.~\eqref{eq:taylor} is at most
of order $\mathcal{O}(\Delta_E^2)$ and can therefore consistently be
neglected with respect to the first one.  It should be noted, however,
that for probability functions which are slowly varying on the bin's
energy range (so that their first-order expansion, $P(\bar{E} + \xi_E)
\approx \bar{P} + P_E'\, \xi_E$, is a good approximation as we assumed
for $N$), then the second term in eq.~\eqref{eq:taylor} is of order
$\mathcal{O}(\Delta_E^3)$ (because $\langle \xi_E / \Delta_E \rangle =
0$, so that the leading $\bar{P}$ contribution vanishes).  This
suggests that keeping only the first term in the expansion,
$N_\text{ch}^i \propto \Delta_E \cdot \bar{N} \cdot \langle P
\rangle$, may result in an $\mathcal{O}(\Delta_E^3)$ approximation at
least in some case.  We will return on this later on.

\subsection{Average over energy}

As we have just seen, calculating the integral in
eq.~\eqref{eq:integral} requires estimating the average value of both
functions $N(E)$ and $P(E)$ over the range of each energy interval.
The former is pretty easy, as under the assumption that the
first-order expansion is accurate enough within the bin, we can simply
use the value of $N$ in the central point of the bin: $\bar{N} =
N(\bar{E})$.  Alternatively, if $N$ does not depend on the parameters
of the model (as it is the case when the physics model under
consideration only affect neutrino propagation), we can afford
estimating $\bar{N} = \langle N \rangle$ numerically by subdividing
the bin into smaller parts, as in any case this is a one-time-only
calculation.

In order to calculate the average probability $\langle P \rangle$, we
take advantage of eq.~\eqref{eq:factorS}.  Let $\alpha$ and $\beta$
denote the initial and final neutrino flavor state, so that $P(E)
\equiv |\bm{S}_{\beta\alpha}(E)|^2$.  Then:
\begin{equation}
  P(\bar{E} + \xi_E) = \Big|
  \big[ \bar{\bm S}\, e^{-i \bm{K}_E\, \xi_E} \big]_{\beta\alpha}
  \Big|^2 = \Big|\big[
    \bar{\bm S} \bm{V}_E\, e^{-i \bm\lambda_E\, \xi_E}\, \bm{V}_E^\dagger
    \big]_{\beta\alpha} \Big|^2
\end{equation}
where we have introduced the matrix $\bm{V}_E$ diagonalizing
$\bm{K}_E$:
\begin{equation}
  \label{eq:avgbasis}
  \bm{V}_E^\dagger\, \bm{K}_E\, \bm{V}_E
  = \bm\lambda_E
  \quad\text{with}\quad
  \bm\lambda_E = \diag\{ \lambda_i^E \} \,.
\end{equation}
Expanding in components:
\begin{multline}
  P(\bar{E} + \xi_E) = \Big| \sum_i
  \big[ \bar{\bm S} \bm{V}_E \big]_{\beta i}\,
  e^{-i \lambda_i^E\, \xi_E}\,
  \big[ \bm{V}_E^\dagger \big]_{i\alpha} \Big|^2
  \\
  = \sum_{ij}
  \big[ \bar{\bm S} \bm{V}_E \big]_{\beta i} \,
  \big[ \bar{\bm S} \bm{V}_E \big]_{\beta j}^* \,
  \big[ \bm{V}_E \big]_{\alpha i}^* \,
  \big[ \bm{V}_E \big]_{\alpha j} \,
  e^{i (\lambda_j^E - \lambda_i^E) \xi_E} \,.
\end{multline}
Averaging $P(E)$ over the bin's energy range reduces to calculating
$\big\langle e^{i (\lambda_j^E - \lambda_i^E) \xi_E} \big\rangle$ for
$\xi_E \in [-\Delta_E/2,\, +\Delta_E/2]$, which gives:
\begin{equation}
  \label{eq:PmeanE}
  \langle P \rangle = \sum_{ij}
  \big[ \bar{\bm S} \bm{V}_E \big]_{\beta i} \,
  \big[ \bar{\bm S} \bm{V}_E \big]_{\beta j}^* \,
  \big[ \bm{V}_E \big]_{\alpha i}^* \,
  \big[ \bm{V}_E \big]_{\alpha j} \,
  \sinc\bigg( \frac{(\lambda_j^E - \lambda_i^E)\, \Delta_E}{2} \bigg) \,.
\end{equation}
It is interesting to notice that for $\Delta_E = 0$ this expression
immediately reduces to $\langle P \rangle = P(\bar{E})$, so a
numerical code which implements averaging as described here can also
trivially provide unaveraged results.  Such situation also arises when
the eigenvalues of $\bm{K}_E$ are ``small'', $(\lambda_j^E -
\lambda_i^E) \Delta_E \ll 1$, in which case the oscillation
probabilities vary slowly over the bin's energy range.  In turn, in
the limit of very fast oscillations, $(\lambda_j^E - \lambda_i^E)
\Delta_E \gg 1$, the $\sinc(x)$ function cancels the off-diagonal
contributions (\textit{i.e.}, $\sinc[(\lambda_j^E - \lambda_i^E)
  \Delta_E / 2] \to \delta_{ij}$), so interference among different
($i\ne j$) effective mass eigenstates is suppressed leading to full
decoherence.

For completeness, we also provide the expression of the higher-order
term $\langle P \cdot \xi_E / \Delta_E \rangle$ which appears in
eq.~\eqref{eq:taylor}:
\begin{equation}
  \bigg\langle P \cdot \frac{\xi_E}{\Delta_E} \bigg\rangle
  = \sum_{ij}
  \big[ \bar{\bm S} \bm{V}_E \big]_{\beta i} \,
  \big[ \bar{\bm S} \bm{V}_E \big]_{\beta j}^* \,
  \big[ \bm{V}_E \big]_{\alpha i}^* \,
  \big[ \bm{V}_E \big]_{\alpha j} \cdot
  \frac{1}{2i}
  \sinc'\bigg( \frac{(\lambda_j^E - \lambda_i^E)\, \Delta_E}{2} \bigg) \,.
\end{equation}
Notice that $\sinc'(x) \sim -x/3$ for $x \to 0$, as can be seen in
eq.~\eqref{eq:sincs}, so in the limit $(\lambda_j^E - \lambda_i^E)
\Delta_E \ll 1$ the expression $\langle P \cdot \xi_E / \Delta_E
\rangle$ is suppressed by one power of $\Delta_E$, making the second
term of eq.~\eqref{eq:taylor} of order $\mathcal{O}(\Delta_E^3)$ -- as
already inferred in the introduction of this chapter.  However, if
$(\lambda_j^E - \lambda_i^E) \Delta_E \sim 1$ such suppression does
not take place, and the corresponding correction --~although still
subleading withe respect to the $\langle P\rangle$ contribution~-- is
simply of order $\mathcal{O}(\Delta_E^2)$.  For simplicity, in the
rest of this note we will neglect this term.

\subsection{Average over trajectory}

The generalization of these results to integrals over multiple
dynamical variables follows the same line.  Let us consider the case
of atmospheric neutrinos described in section~\ref{sec:zenith}.  Now
in addition to the neutrino energy $E$ we should integrate also over
the zenith angle $\Theta$, and the expression of $\bm{S}(E, \Theta)$
is given by eq.~\eqref{eq:factorSatm}.  In order to calculate the
average probability $\langle P \rangle$, we first diagonalize the
perturbation matrices $\bm{K}_E$ and $\bm{K}_\Theta$:
\begin{equation}
  \bm{V}_E^\dagger\, \bm{K}_E\, \bm{V}_E
  = \bm\lambda_E = \diag\{ \lambda_i^E \}
  \quad\text{and}\quad
  \bm{V}_\Theta^\dagger\, \bm{K}_\Theta\, \bm{V}_\Theta
  = \bm\lambda_\Theta = \diag\{ \lambda_i^\Theta \} \,.
\end{equation}
With this, denoting by $\alpha$ and $\beta$ the initial and final
neutrino flavor state, we can write:
\begin{multline}
  P(\bar{E} + \xi_E, \bar\Theta + \xi_\Theta)
  = \Big| \big[ \bar{\bm S}\, e^{-i \bm{K}_E\, \xi_E}\,
    e^{-i \bm{K}_\Theta\, \xi_\Theta} \big]_{\beta\alpha} \Big|^2
  = \Big|\big[ \bar{\bm S}
    \bm{V}_E\, e^{-i \bm\lambda_E\, \xi_E}\, \bm{V}_E^\dagger
    \bm{V}_\Theta\, e^{-i \bm\lambda_\Theta\, \xi_\Theta}\, \bm{V}_\Theta^\dagger
    \big]_{\beta\alpha} \Big|^2
  \\
  = \sum_{ijkl}
  \big[ \bar{\bm S} \bm{V}_E \big]_{\beta i}
  \big[ \bar{\bm S} \bm{V}_E \big]_{\beta j}^*
  \big[ \bm{V}_E^\dagger \bm{V}_\Theta \big]_{ik}
  \big[ \bm{V}_E^\dagger \bm{V}_\Theta \big]_{jl}^*
  \big[ \bm{V}_\Theta^\dagger \big]_{k\alpha}
  \big[ \bm{V}_\Theta^\dagger \big]_{l\alpha}^* \,
  e^{i (\lambda_j^E - \lambda_i^E) \xi_E}\,
  e^{i (\lambda_l^\Theta - \lambda_k^\Theta) \xi_\Theta}
\end{multline}
which, after averaging over their respective bin intervals $\Delta_E$
and $\Delta_\Theta$, yield:
\begin{multline}
  \label{eq:avgPbidi}
  \langle P \rangle = \sum_{ijkl}
  \big[ \bar{\bm S} \bm{V}_E \big]_{\beta i} \,
  \big[ \bar{\bm S} \bm{V}_E \big]_{\beta j}^* \,
  \big[ \bm{V}_E^\dagger \bm{V}_\Theta \big]_{ik} \,
  \big[ \bm{V}_E^\dagger \bm{V}_\Theta \big]_{jl}^* \,
  \big[ \bm{V}_\Theta^\dagger \big]_{k\alpha} \,
  \big[ \bm{V}_\Theta^\dagger \big]_{l\alpha}^* \cdot
  \\
  \sinc\bigg(
  \frac{(\lambda_j^E - \lambda_i^E)\, \Delta_E}{2} \bigg) \,
  \sinc\bigg(
  \frac{(\lambda_l^\Theta - \lambda_k^\Theta)\, \Delta_\Theta}{2}
  \bigg) \,.
\end{multline}
This expression, albeit correct, is not very illuminating.  Things
become clearer if we make use instead of the following algorithm,
which reproduce eq.~\eqref{eq:avgPbidi} by applying a chain of
transformations ($\bm\rho_0 \to \bm\rho_1 \to \bm\rho_2 \to \bm\rho$)
to the density matrix describing the neutrino state:
\begin{enumerate}[label=\emph{\alph*})]
\item we begin by setting the density matrix to the projector onto the
  initial neutrino state:
  \begin{equation}
    \bm\rho_0 \equiv \bm\Pi^{(\alpha)}
    \quad\text{with}\quad
    \bm\Pi^{(\alpha)}_{ij} = \delta_{\alpha i} \delta_{\alpha j} \,;
  \end{equation}

\item we rotate it to the basis where $\bm{K}_\Theta$ is diagonal,
  multiply it element-wise by a matrix $\bm{G}_\Theta$, and rotate it
  back to the flavor basis:
  \begin{equation}
    \label{eq:smearZ}
    \left.
    \begin{aligned}
      \bm\rho_0
      &\to \bm\rho_0' \equiv
      \bm{V}_\Theta^\dagger\, \bm\rho_0\, \bm{V}_\Theta
      \\
      &\to \bm\rho_0'' \equiv
      \bm\rho_0' \odot \bm{G}_\Theta
      \\
      &\to \bm\rho_1 \equiv
      \bm{V}_\Theta\, \bm\rho_0''\, \bm{V}_\Theta^\dagger
    \end{aligned}
    \enspace \right\}
    \quad\text{with}\quad
    \bm{G}^\Theta_{kl} \equiv
    \sinc\bigg(
    \frac{(\lambda_l^\Theta - \lambda_k^\Theta)\, \Delta_\Theta}{2}
    \bigg) \,;
  \end{equation}

\item we rotate it to the basis where $\bm{K}_E$ is diagonal, multiply
  it element-wise by a matrix $\bm{G}_E$, and rotate it back to the
  flavor basis:
  \begin{equation}
    \label{eq:smearE}
    \left.
    \begin{aligned}
      \bm\rho_1
      &\to \bm\rho_1' \equiv
      \bm{V}_E^\dagger\, \bm\rho_1\, \bm{V}_E
      \\
      &\to \bm\rho_1'' \equiv
      \bm\rho_1' \odot \bm{G}_E
      \\
      &\to \bm\rho_2 \equiv
      \bm{V}_E\, \bm\rho_1''\, \bm{V}_E^\dagger
    \end{aligned}
    \enspace \right\}
    \quad\text{with}\quad
    \bm{G}^E_{ij} \equiv
    \sinc\bigg(
    \frac{(\lambda_j^E - \lambda_i^E)\, \Delta_E}{2}
    \bigg) \,;
  \end{equation}

\item we apply the evolution operator $\bar{\bm S}$, thus obtaining
  the density matrix $\bm\rho$ at the detector:
  \begin{equation}
    \bm\rho_2 \to \bm\rho \equiv
    \bar{\bm S}\, \bm\rho_2\, \bar{\bm S}^\dagger \,.
  \end{equation}
\end{enumerate}
The average probability is then given by $\langle P \rangle = \Tr\big[
  \bm\rho \, \bm\Pi^{(\beta)} \big] = \bm\rho_{\beta\beta}$.  In any
case, it should be noticed that this approach involves the
construction of the entire neutrino density matrix, so that it is
readily at hand in situations when the bare probabilities do not
suffice (for example, in the presence of flavor-changing neutrino
interactions in the detector, as is the case for NSI with
electrons~\cite{Coloma:2022umy}).

The sequential approach described above makes it manifest the way
averaging acts.  Each matrix $\bm{K}_X$ associated to a dynamical
variable $X$ gets decomposed into two parts: its eigenvalues, which
induce decoherence by suppressing the off-diagonal elements of the
density matrix, and its diagonalizing matrix $\bm{V}_X$, which
determines in which basis the aforesaid suppression takes place.
Averaging over different variables ($E$ and $\Theta$ in our example)
results in subsequent decoherence applied in different basis.  As
already noted at the beginning of section~\ref{sec:zenith}, the
\emph{order} in which we average over $E$ and $\Theta$ affects the
final result, but only at subleading order $\Delta_E\, \Delta_\Theta$.

From the mathematical point of view, decoherence is introduced through
element-wise multiplication of the density matrix $\bm\rho$ by a
suitable matrix $\bm{G}_X$.  This process does not spoil the
hermiticity of $\bm\rho$ since $\bm{G}_X$ is itself hermitian.
Furthermore, the condition $\Tr(\bm\rho) = 1$ is unaltered as the
diagonal entries of $\bm{G}_X$ are identically $1$ by construction.
Finally, the property $\Tr(\bm\rho^2) \le 1$ is preserved since
$|\bm{G}_{ij}^X| \le 1$ for any $i\ne j$ pair.

\subsection{Integral over production point}
\label{sec:altitude}

Sometimes the neutrino source has a sizable spatial extension, so that
the integration over the production region cannot be neglected.  Such
integral can be formally decomposed into two components: the
\emph{longitudinal} one, corresponding to the direction of neutrino
propagation, and the \emph{transversal} one, which is orthogonal to
it.  Integration over the transversal variables $X$ is just a special
realization of the ``average over trajectory'' discussed in the
previous section, and can therefore be described in terms of suitable
perturbation matrices $\bm{K}_X$ and smearing matrices $\bm{G}_X$.
The same is certainly true also for the longitudinal integral, with
the extra benefit that the corresponding changes to the trajectory
only affect a little portion at its beginning and are therefore
straightforward to implement.  On the other hand, the integral over
the longitudinal direction becomes even simpler when neutrino
propagation exhibits translational invariance inside the production
region, as stated in property~\ref{it:vacuum-L} at the start of
chapter~\ref{cha:taylor}.
A typical example is provided by atmospheric neutrinos, for which the
oscillation probabilities depend not only on the energy and zenith
angle, but also on the altitude of the production point in the
atmosphere.
Usually the air matter density can safely be neglected, so that
propagation proceeds as in vacuum and is described by an Hamiltonian
$\bm{H}_0$ (be it the usual vacuum term, or a different one if New
Physics is present) independent of the position.
In the rest of this section we will focus on this particular case,
using atmospheric neutrinos as a guideline.

Let us begin by fixing the neutrino energy and zenith angle to
reference values $\bar{E}$ and $\bar\Theta$, and neglecting their
variation at first.  In this case, denoting by $\ell$ the slant height
of the production point (\textit{i.e.}, the distance to the ground
level as measured along the neutrino trajectory, which is not
necessarily vertical), we have:
\begin{equation}
  \label{eq:altitude}
  \bm{S}(\ell) = \bm{S}(0)\, e^{-i \bm{H}_0 \ell}
  \quad\Rightarrow\quad
  \bm{S}(\bar\ell + \xi_\ell) = \bar{\bm S}\, e^{-i \bm{H}_0 \xi_\ell}
  \quad\text{with}\quad
  \bar{\bm S} \equiv \bm{S}(\bar\ell)
\end{equation}
where $\xi_\ell$ is the distance from a reference position $\bar\ell$,
which may or may not coincide with ground level.  The first thing to
notice is the formal similarity of this expression with
eq.~\eqref{eq:factorS}, the main conceptual difference being that
eq.~\eqref{eq:altitude} is exact for any $\xi_\ell$ and not the
outcome of a first-order expansion.  Letting $\bm{U}_0$ be the matrix
which diagonalizes $\bm{H}_0$, so that $\bm{U}_0^\dagger\, \bm{H}_0\,
\bm{U}_0 = \bm\omega_0 = \diag\{ \omega_i^0 \}$, we can write:
\begin{equation}
  P(\bar{\ell} + \xi_\ell) = \Big|
  \big[ \bar{\bm S}\, e^{-i \bm{H}_0 \xi_\ell} \big]_{\beta\alpha}
  \Big|^2 = \sum_{ij}
  \big[ \bar{\bm S} \bm{U}_0 \big]_{\beta i} \,
  \big[ \bar{\bm S} \bm{U}_0 \big]_{\beta j}^* \,
  \big[ \bm{U}_0 \big]_{\alpha i}^* \,
  \big[ \bm{U}_0 \big]_{\alpha j} \,
  e^{i (\omega_j^0 - \omega_i^0) \xi_\ell} \,.
\end{equation}
The next step is to average over the altitude of the production point.
Denoting by $\pi_\ell(\xi_\ell)$ the probability density of creating a
neutrino at slant height $(\bar\ell + \xi_\ell)$, we get:
\begin{equation}
  \label{eq:ProbAtmAlt}
  \langle P \rangle = \sum_{ij}
  \big[ \bar{\bm S} \bm{U}_0 \big]_{\beta i} \,
  \big[ \bar{\bm S} \bm{U}_0 \big]_{\beta j}^* \,
  \big[ \bm{U}_0 \big]_{\alpha i}^* \,
  \big[ \bm{U}_0 \big]_{\alpha j} \,
  \int \pi_\ell(\xi_\ell)\,
  e^{i (\omega_j^0 - \omega_i^0) \xi_\ell}\, \dd\xi_\ell \,.
\end{equation}
Hence, thanks to the assumption of translational invariance of
$\bm{H}_0$, the integral over the neutrino production point can be
performed in a single shot, without the need of splitting the
integration domain into smaller steps.  Furthermore,
eq.~\eqref{eq:ProbAtmAlt} is impressively similar to the formalism
presented in the previous sections, which suggests that its numerical
implementation can be easily merged with the average over the neutrino
energy and direction.  Indeed, this is accomplished by modifying the
algorithm in section~\ref{sec:zenith} as follows:
\begin{enumerate}[resume*, start=4]
\item[\emph{a--c})] we proceed as before until the construction of the
  density matrix $\bm\rho_2$;

\item we rotate it to the basis where $\bm{H}_0$ is diagonal, multiply
  it element-wise by a matrix $\bm{G}_\ell$, and rotate it back to the
  flavor basis:
  \begin{equation}
    \label{eq:smearA}
    \left.
    \begin{aligned}
      \bm\rho_2
      &\to \bm\rho_2' \equiv
      \bm{U}_0^\dagger\, \bm\rho_2\, \bm{U}_0
      \\
      &\to \bm\rho_2'' \equiv
      \bm\rho_2' \odot \bm{G}_\ell
      \\
      &\to \bm\rho_3 \equiv
      \bm{U}_0\, \bm\rho_2''\, \bm{U}_0^\dagger
    \end{aligned}
    \enspace \right\}
    \quad\text{with}\quad
    \bm{G}^\ell_{ij} \equiv
    \int \pi_\ell(\xi_\ell)\,
    e^{i (\omega_j^0 - \omega_i^0) \xi_\ell}\, \dd\xi_\ell \,;
  \end{equation}

\item we apply the evolution operator $\bar{\bm S}$, thus obtaining
  the density matrix $\bm\rho$ at the detector:
  \begin{equation}
    \bm\rho_3 \to \bm\rho \equiv
    \bar{\bm S}\, \bm\rho_3\, \bar{\bm S}^\dagger \,.
  \end{equation}
\end{enumerate}
As can be seen, this approach treats averaging over neutrino energy,
arrival direction and production altitude on the same footing.  It
should be noted, however, that our formalism is based on a first-order
expansion, and therefore relies on the assumption that the three
dynamical variables $\xi_E$, $\xi_\Theta$ and $\xi_\ell$ are
sufficiently small.  While $\xi_E$ and $\xi_\Theta$ can be kept under
control by suitably choosing their corresponding intervals $\Delta_E$
and $\Delta_\Theta$, the range of $\xi_\ell$ is determined by the
properties of the Earth's atmosphere (or more in general of the
neutrino source), and cannot be changed.  Yet this does not spoil the
accuracy of the calculations when $\Delta_E = \Delta_\Theta = 0$ as
long as translational invariance ensures that eq.~\eqref{eq:altitude}
is exact.  In other words, our procedure may fail to account for terms
of order $\xi_E\, \xi_\ell$ or $\xi_\Theta\, \xi_\ell$ when deviating
from the reference ray, but while the smallness of $\xi_\ell$ is not a
priori guaranteed for arbitrary sources, such terms are still
subleading due to the smallness of $\xi_E$ and $\xi_\Theta$.  This
issue can be further mitigated by tuning the reference altitude
$\bar\ell$, for example ensuring that the mean of $\pi_\ell(\xi_\ell)$
is zero.  Anyway, the size of the Earth's atmosphere is indeed small
compared to the overall radius of the Earth, hence in this case the
validity of the computation is secured by the physical system.  And of
course, for non-uniform sources (such as the core of the Sun) or when
the interplay between the source's overall extension and the size of
the energy and zenith integration bins cannot be neglected, one always
have the option of splitting the production range into steps small
enough to circumvent these issues, and handle the longitudinal
integral numerically.

Looking at the definition of $\bm{G}_\ell$ in eq.~\eqref{eq:smearA},
we see that its elements are strictly related to the Fourier transform
$\hat\pi_\ell$ of the altitude distribution function $\pi_\ell$:
$\bm{G}_{ij}^\ell \equiv \hat\pi_\ell(\omega_i^0 - \omega_j^0)$.  This
is also the case for $\bm{G}_E$ or $\bm{G}_\Theta$, since they were
constructed assuming uniform priors within their respective ranges:
$\pi_E(\xi_E) \equiv \rect(\xi_E / \Delta_E) / \Delta_E$ and similarly
for $\pi_\Theta(\xi_\Theta)$, whose Fourier transform is indeed the
$\sinc(x)$ function.  This suggests that the flat averaging over the
bin's range which we have performed so far can be generalized by
assuming alternative distributions for $\xi_E$ and $\xi_\Theta$.  For
example, a Gaussian prior for $\pi_E(\xi_E)$ would yield:
\begin{equation}
  \pi_E(\xi_E) = \frac{1}{\sqrt{2\pi}\,\Delta_E} \exp\bigg\{
    {-\frac{1}{2}} \bigg[ \frac{\xi_E}{\Delta_E} \bigg]^2 \bigg\}
  \quad\Rightarrow\quad
  \bm{G}_{ij}^E = \exp\bigg\{ {-\frac{1}{2}}
  \big[ (\lambda_i^E - \lambda_j^E)\, \Delta_E \big]^2 \bigg\} \,.
\end{equation}
This is precisely the idea behind the low-pass filter in
ref.~\cite{Huber:2007ji}, and can be useful to describe,
\textit{e.g.}, the smearing of the oscillation probabilities induced
by a finite energy resolution $\Delta_E$ of the detector -- provided
that $\Delta_E$ is small enough for our first-order expansion to hold.
Alternatively, in Monte-Carlo calculations where the dynamical
variables $E$ And $\Theta$ are chosen randomly by an integrator
routine and no energy or angular grid are defined, it may be
convenient to introduce exponential smearing on scales $\Delta_E$ and
$\Delta_\Theta$ well below the resolution of the detector, so to
properly handle fast oscillations without spoiling the reliability of
the simulation.

\subsection{Tabulation and interpolation}

To conclude this chapter, let's briefly comment on a trivial extension
of the techniques described so far.  In eq.~\eqref{eq:factorSatm} we
have illustrated how the perturbation matrices $(\bm{K}_E,
\bm{K}_\Theta)$ can be used to ``shift'' the evolution matrix $\bm{S}$
from its central value $\bar{\bm S}$ calculated at $(\bar{E},
\bar\Theta)$ to a nearby position $(\bar{E} + \xi_E, \bar\Theta +
\xi_\Theta)$.  In the previous sections we have used this formula to
derive accurate averages over energy and zenith angle, assuming some
distribution $\pi_E(\xi_E)$ and $\pi_\Theta(\xi_\Theta)$ (either plain
rectangular functions with widths $\Delta_E$ and $\Delta_\Theta$, or
more general ones such as Gaussian priors) around the central value
$(\bar{E}, \bar\Theta)$.  However, our formalism trivially allows to
perform averages also around \emph{shifted} values, $(\bar{E} +
\delta_E, \bar\Theta + \delta_\Theta)$.  This is accomplished by means
of shifted priors, $\pi_E(\xi_E - \delta_E)$ and
$\pi_\Theta(\xi_\Theta - \delta_\Theta)$, which leads to a rephasing
of the $\bm{G}_E$ and $\bm{G}_\Theta$ matrices:
\begin{equation}
  \bm{G}_{ij}^E
  \to e^{i (\lambda_j^E - \lambda_i^E) \delta_E}\, \bm{G}_{ij}^E
  \quad\text{and}\quad
  \bm{G}_{ij}^\Theta
  \to e^{i (\lambda_j^\Theta - \lambda_i^\Theta) \delta_\Theta}\, \bm{G}_{ij}^\Theta \,.
\end{equation}
This simple observation opens the door to efficient tabulation and
interpolation of oscillation amplitudes.  Consider the case where a
Monte-Carlo generator needs to compute the neutrino conversion
probabilities for a very large number of $(E, \Theta)$ rays.  A
well-known technique to speed up computations is to first tabulate the
probabilities on a representative grid of $(\bar{E}_i, \bar\Theta_j)$
values, and then extract the actual $(E, \Theta)$ ray by
interpolation.  The problem in doing so, however, is that in the
presence of fast oscillations a fixed $(\bar{E}_i, \bar\Theta_j)$ grid
may fail to reproduce the oscillation pattern accurately enough.  The
solution is to tabulate instead the $(\bar{\bm S}, \bm{K}_E,
\bm{K}_\Theta)$ matrices (which for further convenience can also be
factorized at this stage into unitary $(\bar{\bm U}, \bm{V}_E,
\bm{V}_\Theta)$ and diagonal $(\bar{\bm\omega}, \bm\lambda_E,
\bm\lambda_\Theta)$ components) for each $(\bar{E}_i, \bar\Theta_j)$
node, and later use this information to reconstruct the probabilities
once the required $(E, \Theta)$ value is known.  The most
straightforward way to perform this last step is to find the closest
$(\bar{E}_i, \bar\Theta_j)$ node and use it for extrapolation.  A more
refined approach is to locate the $[\bar{E}_i, \bar{E}_{i+1}] \times
[\bar\Theta_j, \bar\Theta_{j+1}]$ cell containing $(E, \Theta)$,
derive an estimate of the conversion probabilities from each of its
vertices, and then produce a weighted average of such estimates as in
ordinary interpolation.  In this second case we can also evaluate the
reliability of the result by comparing the probabilities obtained from
the various vertices, as for accurate calculations they should all be
similar among them.

\section{Examples}
\label{cha:examples}

In this chapter we will present a number of examples to illustrate the
main features of the formalism just introduced.  Concretely, we will
focus on three aspects: Taylor expansion in energy and trajectory
(described in sections~\ref{sec:onelayer}, \ref{sec:multilay} and
\ref{sec:zenith}), improving the accuracy of $\bar{\bm S}$ within a
definite layer (described in section~\ref{sec:improved}), and
averaging in the presence of fast oscillations (described in
chapter~\ref{cha:average}).

\subsection{Taylor expansion in energy and trajectory}

\begin{figure}
 \includegraphics[width=0.98\textwidth]{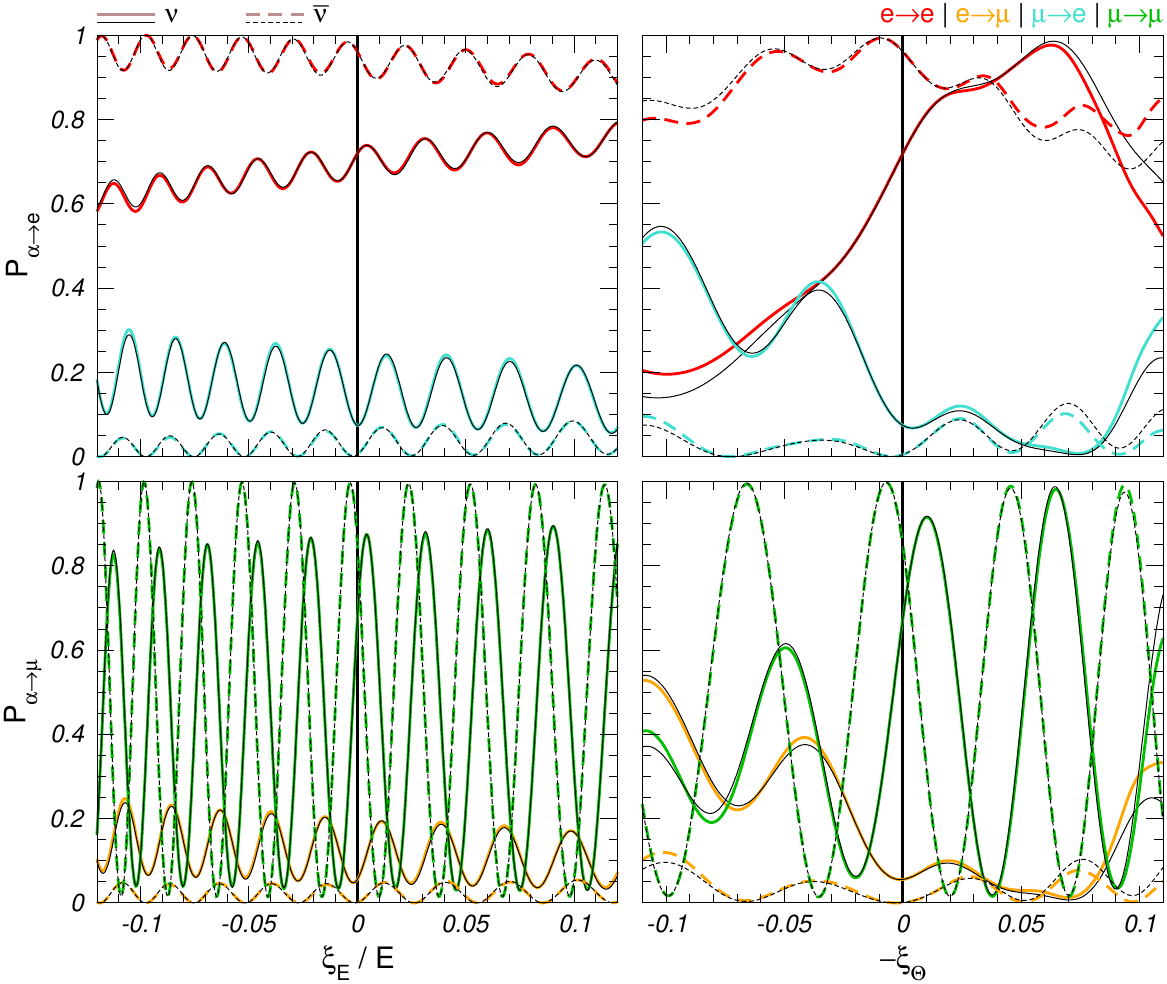}
 \caption{Comparison between exact (colored) and first-order (black)
   probabilities, for neutrinos (solid) and antineutrinos (dashed).
   We assume standard $3\nu$ oscillations with $\sin^2\theta_{12} =
   0.303$, $\sin^2\theta_{13} = 0.022$, $\sin^2\theta_{23} = 0.572$,
   $\Dmq_{21} = 7.41\times 10^{-5}~\eVq$, $\Dmq_{21} = 2.51\times
   10^{-3}~\eVq$ and $\delta_\text{CP} = 197^\circ$.  As reference ray
   we fix $\bar{E} = 0.3$~GeV, $\cos\bar\Theta = -0.9$ and
   $\text{production altitude} = 25~\text{km}$.  In the left (right)
   panel we show the effects of modifying the energy (direction).}
 \label{fig:taylor-A}
\end{figure}

\begin{figure}
 \includegraphics[width=0.98\textwidth]{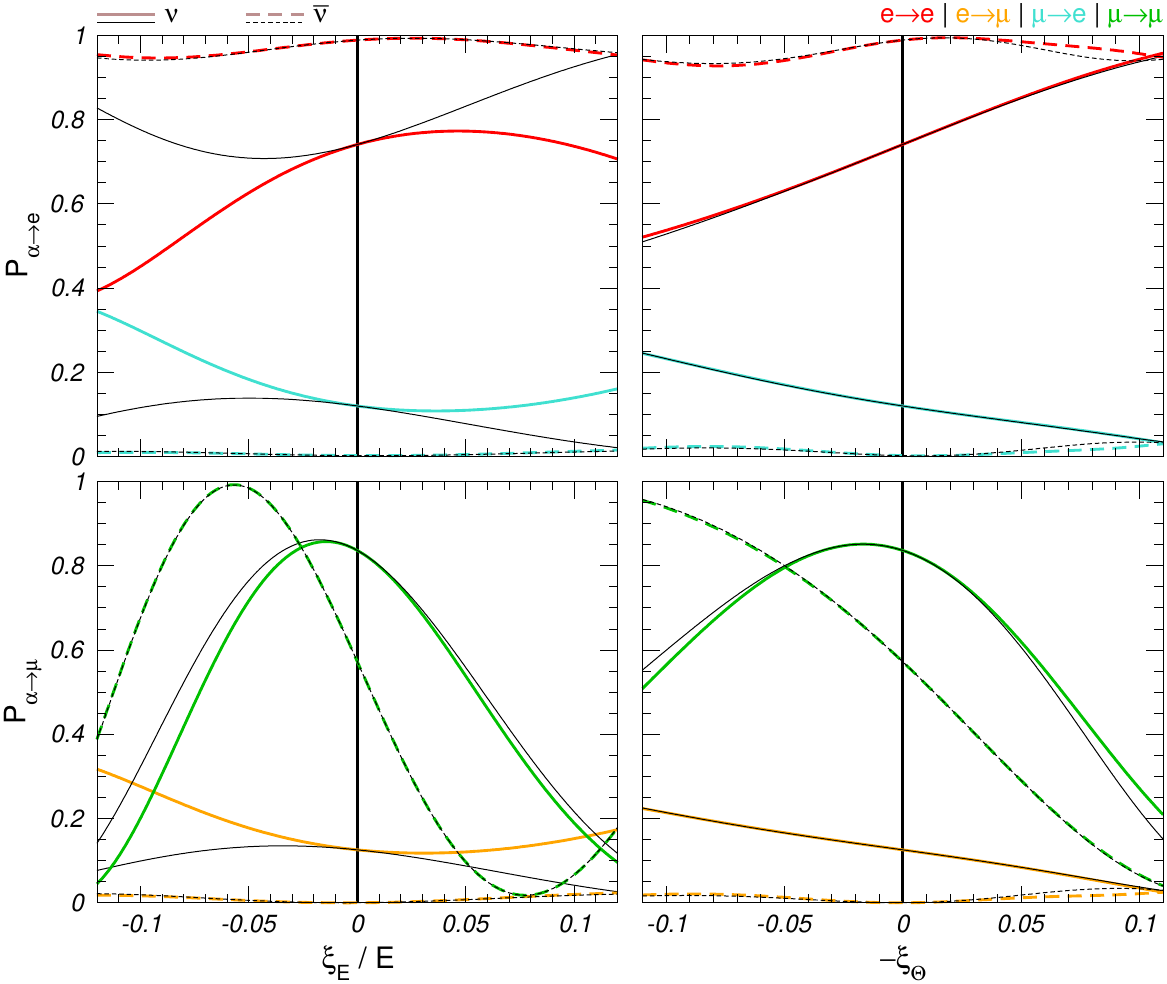}
 \caption{Same as figure~\ref{fig:taylor-A} but for $\bar{E} = 3$~GeV.
   In the left (right) panel we show the effects of modifying the
   energy (direction).  See text for details.}
 \label{fig:taylor-B}
\end{figure}

In figures~\ref{fig:taylor-A} and~\ref{fig:taylor-B} we plot the
oscillation probabilities in various channels for atmospheric
neutrinos (solid lines) and antineutrinos (dashed lines) crossing the
Earth matter.  We assume standard three-neutrino oscillations and set
the corresponding parameters to the NuFIT-5.2 best-fit
value~\cite{Esteban:2020cvm, nufit-5.2}.  We fix $\bar{E} = 0.3$~GeV
(figure~\ref{fig:taylor-A}) or $\bar{E} = 3$~GeV
(figure~\ref{fig:taylor-B}) as reference value for the neutrino
energy, as well as $\cos\bar\Theta = -0.9$ for the zenith angle of the
arrival direction, and compute the matrices $(\bar{\bm S}, \bm{K}_E,
\bm{K}_\Theta)$ defined in chapter~\ref{cha:taylor}.  We then plot the
dependence of the probabilities on the neutrino energy $E = \bar{E} +
\xi_E$ (left panels) and zenith angle $\Theta = \bar\Theta +
\xi_\Theta$ (right panels), and compare the exact calculations (thick
colored lines) with the extrapolation based on
eq.~\eqref{eq:factorSatm} (thin black lines).

As can be seen, all black lines in all panels match the value of their
colored counterpart at zero shift.  This is by construction, as that
corresponds precisely to the reference value used for the calculation
of $\bar{\bm S} \equiv \bm{S}(\bar{E}, \bar\Theta)$.  However, in
addition to point-like coincidence the black lines are also
\emph{tangent} to the colored ones, and this is a consequence of
taking into account also the first-order terms.  To make it clear, if
we had neglected $\bm{K}_E$ and $\bm{K}_\Theta$ in our calculations
--~thus sticking just to the usual zero-order approximation~-- all the
black lines would have been perfectly horizontal.

Another feature of our correction terms is that they capture the
relevant oscillation ``frequencies'' (in $E$ and $\Theta$) of the
system, so that the black lines can often ``track'' the exact
calculations for a sizable interval around zero.  This is particularly
the case at low energy (see figure~\ref{fig:taylor-A}), when
oscillations are dominated by the vacuum term for which our formalism
becomes exact.  Notice, however, that even at $\bar{E} = 0.3$~GeV
matter effects still play an important role, as the clear differences
between same-channel neutrino and antineutrino probabilities
demonstrate, yet this does not spoil the accuracy of the
extrapolation.

On the other hand, figure~\ref{fig:taylor-B} illustrates what our
approximation \emph{cannot} do.  At $\bar{E} = 3~\text{GeV}$ the
interference effects between the vacuum and matter terms of the
Hamiltonian are pretty strong.  As a first-order expansion, the
matrices $\bm{K}_E$ and $\bm{K}_\Theta$ catch very little of the
non-commutativity of the system, so they cannot help in estimating the
``curvature'' of the lines (which is a second-order effect) beyond the
simple periodic oscillation pattern.  This is clearly visible in the
red, orange and cyan neutrino lines, where our extrapolation sizably
deviates from the exact result already for energy shifts at the
few-percent level.  This underlines that our method is \emph{not}
intended for large-scale extrapolations, and in particular the size of
the integration bin should be kept small enough to ensure that any
non-oscillatory effect is properly accounted for numerically.  In
other words, the formalism described here takes care of potentially
\emph{fast} oscillations stemming from large derivatives of the
evolution Hamiltonian, but the features of the \emph{slow} oscillation
pattern still require a dense grid in the $(E, \Theta)$ plane.

\subsection{Improved in-layer calculation}

\begin{figure}
 \includegraphics[width=0.98\textwidth]{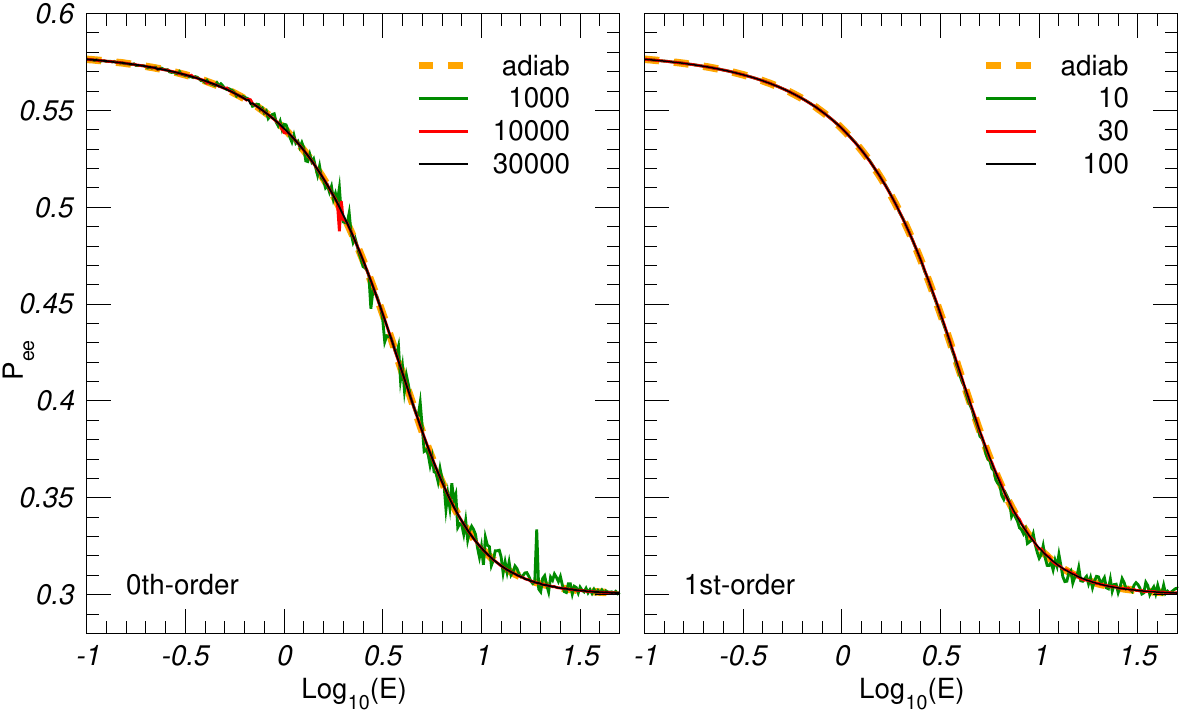}
 \caption{Asymptotic survival probability for a neutrino produced in
   the center of the Sun.  We assume a $2\nu$ oscillation model with
   $\sin^2\theta = 0.3$ and $\Dmq = 7.4\times 10^{-5}~\eVq$.  In both
   panels the orange dashed line is obtained with the adiabatic
   formula.  The colored solid lines correspond to fully numerical
   calculations, with the trajectory inside the Sun divided into as
   many layers as indicated in the legend.  In the left panel we
   assume that the matter density is constant within each layer, while
   in the right panel we account for the first-order correction
   described in section~\ref{sec:improved}.}
 \label{fig:inlayer}
\end{figure}

Neutrino propagation in arbitrary matter profiles can be handled by
dividing the path into a number of sufficiently small layers.  As
described in section~\ref{sec:improved}, within each layer we can
either assume plain constant density, or add a correction proportional
to the first derivative of the matter potential.  To illustrate the
advantage of the second choice, in figure~\ref{fig:inlayer} we plot
the $P_{ee}$ survival probability for a neutrino produced in the
center of the Sun and detected at infinite distance.  For definiteness
we assume two-neutrino oscillations with $\sin^2\theta = 0.3$ and
$\Dmq = 7.4\times 10^{-5}~\eVq$, and the solar matter distribution and
chemical composition given in ref.~\cite{Vinyoles:2016djt}.  For such
a model the MSW effect takes place~\cite{Wolfenstein:1977ue,
  Mikheyev:1985zog} and neutrino probabilities can be calculated
analytically using the adiabatic approximation.  This result is an
excellent benchmark to check the accuracy of our formalism, therefore
we have plotted it in both panels as a thick dashed orange line.

The solid lines in figure~\ref{fig:inlayer} have been computed in a
fully numerical way.  Concretely, we have divided the trajectory
inside the Sun into as many layers as indicated in the legend, and we
have obtained the overall evolution matrix $\bar{\bm S}$ for the full
path by multiplying together the contribution of the various layers.
In the left panel we have assumed constant density within each layer,
as described in eq.~\eqref{eq:Slayer0}.  As can be seen, in order for
the numerical calculation to reproduce the analytic result accurately
enough over the relevant energy range, one need at least
$\mathcal{O}(10^4)$ layers.  Qualitatively, this can be understood as
follows.  A requirement for MSW conversion is that the vacuum and
matter term of the evolution Hamiltonian become comparable at some
point along the trajectory.  These terms can be conveniently
quantified through the oscillation length they induce, $l_\text{osc} =
2\pi / (\bar\omega_2 - \bar\omega_1)$ where $\bar\omega_i$ are the
eigenvalues of the Hamiltonian (see eq.~\eqref{eq:massbasis}).  For
$\Dmq = 7.4\times 10^{-5}~\eVq$ we get $l_\text{osc}^\text{vac} =
33~\text{km}\cdot E/\text{MeV}$ in vacuum, while for $E \to \infty$ we
have $l_\text{osc}^\text{mat} \ge 160~\text{km}$ in solar matter.  The
condition $l_\text{osc}^\text{vac} \sim l_\text{osc}^\text{mat}$
requires $E \gtrsim \text{few MeV}$ (and indeed the transition between
the vacuum-dominated and matter-dominated regime occur in this range,
as clearly visible in figure~\ref{fig:inlayer}) and implies
oscillation lengths $l_\text{osc} \gtrsim \mathcal{O}(100~\text{km})$.
The numerical computation is accurate when the layer size does not
exceed the oscillation length, and this is ensured in all the MSW
region only if the layer length is smaller than
$\mathcal{O}(100~\text{km})$: hence, the number of layers should be at
least $\mathcal{O}(R_\odot / 100~\text{km}) \simeq \mathcal{O}(10^4)$,
with $R_\odot = 7\times 10^5$~km being the solar radius.  Empirically,
one may say that the break of adiabaticity induced by the ``jumps'' in
the potential among consecutive constant-density layers should occur
at scales well below the oscillation length.

On the other hand, in the right panel of figure~\ref{fig:inlayer} we
have taken into account the linear variation of the matter potential
inside each layer, as encoded in eq.~\eqref{eq:Slayer1}.  This
effectively removes the artificial ``jumps'' introduced by the
ladder-like schematization of the potential in the constant-density
limit, and leads to impressively accurate results with as little as a
few tens of layers.  It should be remembered, however, that our
formula for the evolution in a linearly-varying potential is
\emph{not} exact (unlike the constant-density case) but rather
obtained through a perturbative expansion, so the layer length should
always be kept sufficiently small for the approximation to hold.

\subsection{Averaging fast oscillations}

\begin{figure}
 \includegraphics[width=0.98\textwidth]{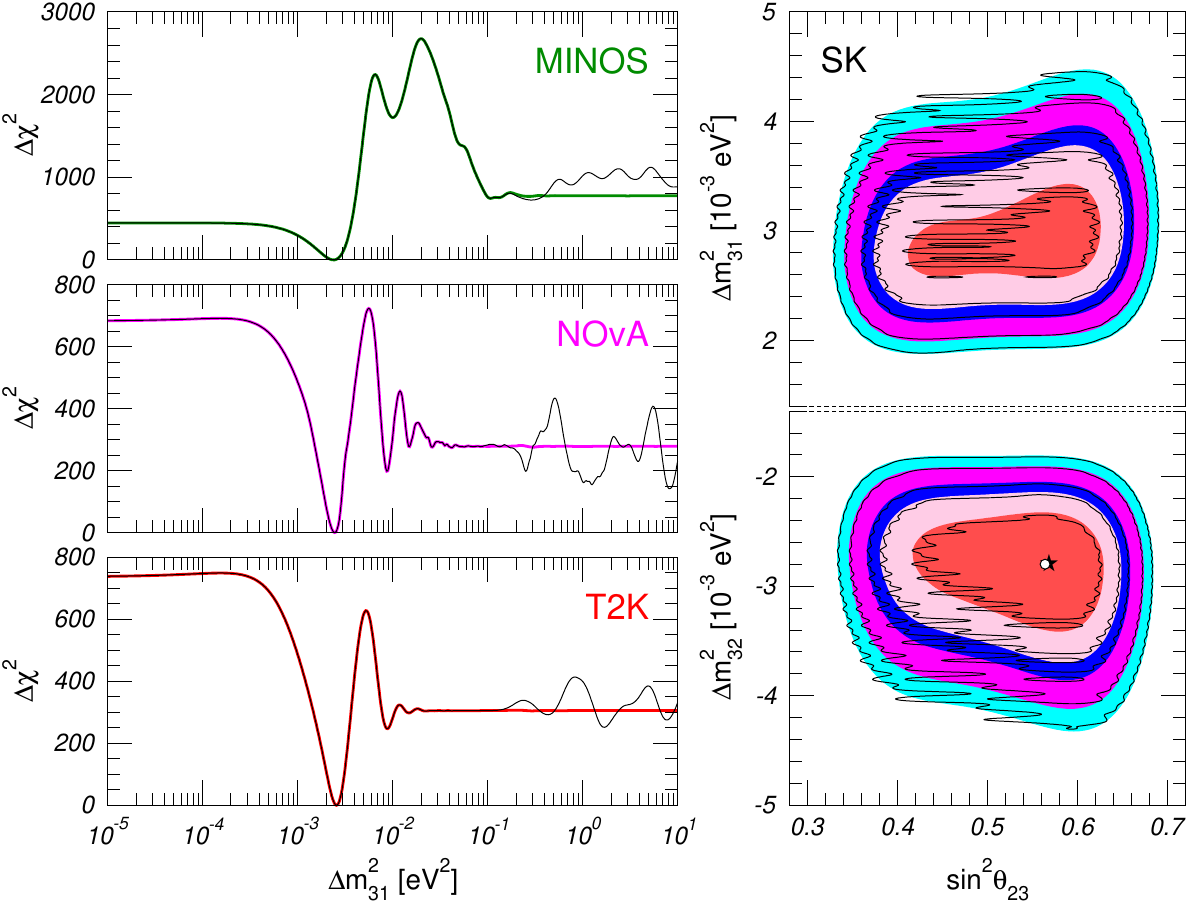}
 \caption{Impact of averaging.  We assume standard $3\nu$ oscillations
   with $\sin^2\theta_{12} = 0.303$, $\sin^2\theta_{13} = 0.022$,
   $\Dmq_{21} = 7.41\times 10^{-5}~\eVq$ and $\delta_\text{CP} =
   197^\circ$.  In the left panels we fix $\sin^2\theta_{23} = 0.572$
   and plot $\Delta\chi^2$ as a function of $\Dmq_{31}$ for various
   accelerator neutrino experiments.  In the right panel we show the
   allowed regions (at $1\sigma$, $90\%$, $2\sigma$, $99\%$, $3\sigma$
   CL for 2 d.o.f.) in the $(\theta_{23}, \Dmq_{3\ell})$ plane from
   the global analysis of Super-Kamiokande atmospheric data.  The
   colored lines or regions are based on the averaging procedure
   described in chapter~\ref{cha:average}, while the black lines do not.
   See text for details.}
 \label{fig:average}
\end{figure}

In chapter~\ref{cha:average} we have shown how the perturbation matrix
$\bm{K}_E$ (and $\bm{K}_\Theta$ for atmospheric neutrinos) can be used
to improve the calculation of the energy integral (and zenith-angle
one) commonly required to estimate the theoretical prediction of a
given measurement.  In particular, our approach naturally handles fast
neutrino oscillations, yielding properly averaged results without the
need of ad-hoc solutions.  To illustrate this feature, in
figure~\ref{fig:average} we consider various experiments and compare
the fits obtained with and without the inclusion of $\bm{K}_E$ and
$\bm{K}_\Theta$.  For definiteness we assume standard $3\nu$
oscillations and fix the undisplayed parameters to the NuFIT-5.2
best-fit value~\cite{Esteban:2020cvm, nufit-5.2}.

In the left panels we focus on accelerator experiments
MINOS~\cite{MINOS:2014rjg}, NOvA~\cite{NOvA:2021nfi} and
T2K~\cite{T2K:2023smv} and plot the overall $\Delta\chi^2$ (defined
with respect to the local minimum) as a function of $\Dmq_{31}$.  The
energy integral for each experiment is converted into a sum by
subdividing the relevant range into uniform bins in logarithmic scale,
with a density of 100 bins per decade.  The central point of each
energy bin is chosen as representative value for the entire bin and
used to calculate neutrino propagation, encoded in the evolution
matrix $\bar{\bm S}$.  This simple integration method is
straightforward to implement, and produce accurate results until well
beyond the boundaries of the experimentally allowed region.  However,
for the sake of illustration we are interested here in the domain
$\Dmq_{31} \gtrsim 10^{-1}~\eVq$, even though it is completely ruled
out by the data.  In this limit oscillations become so fast that the
conversion probabilities can no longer be regarded as ``constant''
within an energy bin.  If this fact is ignored and the probabilities
are still extracted solely from the bin's representative $\bar{\bm S}$
matrix, as is the case for the black lines in
figure~\ref{fig:average}, then the calculation becomes unreliable due
to aliasing effects.  Conversely, if the perturbation matrix
$\bm{K}_E$ is also taken into account as described in
eqs.~\eqref{eq:avgbasis} and~\eqref{eq:PmeanE}, then fast oscillations
are automatically averaged and accuracy is recovered.

As a further example, in the right panel of figure~\ref{fig:average}
we plot the allowed region in the $(\theta_{23}, \Dmq_{3\ell})$ plane
from our own global analysis of Super-Kamiokande atmospheric
data~\cite{Wendell:2014dka}.  In this case the energy integral is
estimated with a density of 50 points per decade in logarithmic scale,
while the neutrino arrival direction is discretized into 100 points
uniformly distributed in $\cos\Theta \in [-1, +1]$.  Despite the very
large number of sampled points (more than $30\,000$ rays in the full
$(E,\Theta)$ plane) the results obtained with calculations based
exclusively on $\bar{\bm S}$ are inaccurate, as illustrated by the
black lines.  This is driven by sub-GeV data, for which $E \lesssim
1~\text{GeV}$ so that the oscillation probabilities of neutrinos
coming from below the horizon are ``fast'' (\textit{i.e.}, they vary a
lot even for small energy and zenith changes).  Once again, taking
into account the information encoded in the perturbation matrices
$\bm{K}_E$ and $\bm{K}_\Theta$ fixes the issue, as can be deduced from
the colored regions.

Of course, one of the reasons behind the failure of calculations based
solely on $\bar{\bm S}$ is that we have chosen a \emph{regular} grid
of sampling points in both energy and zenith angle, which favors
aliasing effects: randomizing our grid would have mitigated the
problem.  Also, for atmospheric neutrinos we have verified that
doubling the density of points (100 per decade in energy, and 200
overall in zenith) significantly improves the quality of the fit (at
least in the standard $3\nu$ case), at the cost of a factor $\sim 4$
in computer time.  In general, various methods exists to handle fast
oscillations, but they all have some kind of drawback.  For example,
one can use an adaptive integration routine which ``detects'' poor
accuracy and adds extra points to compensate for it, but this usually
implies substantial slow-down in difficult regions.  Alternatively,
one can introduce a ``low-pass filter'' as described in
ref.~\cite{Huber:2007ji}, but this requires to choose an ``averaging
length'' according to the details of the experiment under
consideration, furthermore its implementation is only feasible in
limited situations (\textit{e.g.}, in ref.~\cite{Huber:2007ji} this
option is provided just for constant density).  Integration over a
spatially uniform production region can be performed analytically, as
implemented in ref.~\cite{Arguelles:2021twb} and discussed in detail
in section~\ref{sec:altitude}, but it is only effective for averaging
purposes when the oscillation length is smaller than the source's
size, which is not very often unless the source occupies a
considerable fraction of the overall baseline.  Finally, in specific
scenarios where fast oscillations are \emph{known} to occur (such as
in solar neutrinos due to large $\Dmq_{31}$, in atmospheric neutrinos
at sub-GeV energy, or in long-baseline experiments when extra eV-scale
sterile states are considered), it may be possible to factor them out
analytically while leaving the treatment of non-fast oscillations
numerical~(see, \textit{e.g.}, appendices~C and~D of
ref.~\cite{Kopp:2013vaa}) or
semi-analytical~\cite{Ioannisian:2020isl}, but this approach requires
the derivation of appropriate formulas for each propagation model, and
it relies on oscillation frequencies being ``infinitely large'' so
that it cannot handle smooth transitions between ``fast'' and ``slow''
oscillations.  In contrast, our method generically works for any
oscillation model which can be described in terms of an evolution
Hamiltonian, requires a fixed amount of computation time irrespective
of the specific point in parameter space being simulated
(\textit{i.e.}, it is not affected by whether fast oscillations arise
or not), and does not require the choice of an ``averaging length''
because the finite extension of the integration bins take care of that
(and the result is independent of it, as long as the bin is small
enough for the first-order approximation to hold).

\section{Summary}
\label{cha:summary}

In this note we have presented a general formalism which allows to
considerably enhance the accuracy and performance of numerical
neutrino codes needed to calculate the theoretical predictions for
experimentally measured quantities.  In particular:
\begin{itemize}
\item our approach does not make any assumption on the underlying
  theory determining neutrino propagation, hence it can be applied to
  a vast set of models such as standard three-neutrino oscillations,
  extra sterile neutrinos, non-standard neutrino-matter interaction,
  violation of fundamental symmetries, and so on.  Furthermore, it
  works for arbitrary matter density profiles;

\item our method relies on a first-order Taylor expansion of the
  neutrino evolution matrix $\bm{S}(E, \Theta)$ around a reference
  energy $\bar{E}$ and trajectory $\bar\Theta$.  As described in
  eq.~\eqref{eq:factorSatm}, $\bm{S}(\bar{E} + \xi_E, \bar\Theta +
  \xi_\Theta)$ is related to its central value $\bar{\bm S} \equiv
  \bm{S}(\bar{E}, \bar\Theta)$ through suitable perturbation matrices
  $(\bm{K}_E, \bm{K}_\Theta)$, and its unitarity is guaranteed for any
  value of $(\xi_E, \xi_\Theta)$.  The set of $(\bar{\bm S}, \bm{K}_E,
  \bm{K}_\Theta)$ tuples with the multiplication rule in
  eq.~\eqref{eq:tuplemult} forms a group, and provides the building
  blocks to compute neutrino propagation on trajectories comprising
  multiple density layers;

\item our formalism ensures that the integrals over neutrino energy
  and trajectory embedded in the theoretical predictions of
  experimental measurements remain accurate even in the presence of
  fast neutrino oscillations, avoiding aliasing without the need to
  increase the density of integration bins.  This is achieved through
  element-wise multiplication of the neutrino density matrix with
  smearing matrices $(\bm{G}_E, \bm{G}_\Theta)$ in a way entirely
  controlled by the perturbation matrices $(\bm{K}_E, \bm{K}_\Theta)$,
  as seen in eqs.~\eqref{eq:smearZ} and~\eqref{eq:smearE}.  In Riemann
  integration the bin's width naturally acts as a low-pass filter, yet
  its specific value does not affect the final result as long as it is
  small enough for the first-order approximation to hold.  In
  Monte-Carlo simulations a suitable cutoff can be introduced by
  appropriate priors $\pi_E(\xi_E)$ and $\pi_\Theta(\xi_\Theta)$ such
  as Gaussian functions;

\item our method also allows for efficient tabulation and
  interpolation of oscillation amplitudes.  The $(\bar{\bm S},
  \bm{K}_E, \bm{K}_\Theta)$ matrices can be pre-computed on a
  representative grid of $(\bar{E}_i, \bar\Theta_j)$ values, which are
  then used to reconstruct the conversion probabilities once the
  required $(E, \Theta)$ ray is known.  This procedure avoids the loss
  of accuracy and aliasing effects which usually appear in the
  presence of fast oscillations when the probabilities themselves are
  tabulated and interpolated;

\item finally, for atmospheric neutrinos (and in general extended
  sources) our approach unifies averaging over neutrino energy and
  direction with the integral over the neutrino production point, also
  described by a suitable smearing matrix $\bm{G}_\ell$ as shown in
  eq.~\eqref{eq:smearA}.  Furthermore, it naturally leads to the
  construction of the density matrix at the detector, which is
  convenient when considering scenarios where the plain oscillation
  probabilities do not suffice.
\end{itemize}
On the technical side, a pre-existing object-oriented code accounting
for neutrino propagation solely in terms of $\bar{\bm S}$ can be
adapted to incorporate our formalism by replacing the matrix $\bar{\bm
  S}$ and its product with the tuple $(\bar{\bm S}, \bm{K}_E,
\bm{K}_\Theta)$ and the multiplication rule in
eq.~\eqref{eq:tuplemult}.  The addition of the first-order terms does
not result in significant slow-down, as long as the computation time
required for matrix exponentiation is comparable to that of matrix
diagonalization and overwhelms that of matrix multiplication.
Concretely, one extra diagonalization for each $\bm{K}_E$ or
$\bm{K}_\Theta$ matrix is required to perform averaging, which results
in doubling the computation time for constant-density paths (such as
accelerator neutrinos) but negligible impact on trajectories with a
large number of different layers (such as atmospheric neutrinos).
In brief, our approach provides a lossless enhancement with respect to
computations based on $\bar{\bm S}$ alone, which can thus be regarded
as its zero-order limit.

\acknowledgments

We are grateful to P.~Coloma, P.~Denton, M.C.~Gonzalez-Garcia,
E.~Fernandez-Martinez and T.~Ota for useful discussions.
This project is funded by the European Union through the Horizon 2020
research and innovation program (Marie Sk{\l}odowska-Curie grant
agreement 860881-HIDDeN) and the Horizon Europe programme (Marie
Sk{\l}odowska-Curie Staff Exchange grant agreement
101086085-ASYMMETRY).  It also receives support from Spanish grants
PID2019-110058GB-C21 and IFT ``Centro de Excelencia Severo Ochoa''
CEX2020-001007-S funded by MCIN/AEI/10.13039/501100011033.

\appendix

\section{Exploiting the symmetries of the system}
\label{app:splitting}

Let us consider a system described by an Hamiltonian $\mathring{\bm
  H}(E) \equiv \bm{O}\,\bm{H}(E)\, \bm{O}^\dagger$ where $\bm{O}$ is a
unitary matrix.  This situation occur, for example, in standard
three-neutrino oscillations, where the matrix $\bm{O}$ accounts for
the $\theta_{23}$ and $\delta_\text{CP}$ parameters while the
``reduced'' Hamiltonian $\bm{H}(E)$ depends solely on $\theta_{12}$,
$\theta_{13}$, $\Dmq_{21}$, $\Dmq_{31}$.  In this case, it is
immediate to see that $\mathring{\bm S}(E) \equiv \bm{O}\,\bm{S}(E)\,
\bm{O}^\dagger$ for all energies, so that one can perform the bulk of
calculations in the so-called ``propagation basis'' using the reduced
matrix $\bm{H}(E)$ (which is simpler and depends on less parameters)
and then reintroduce $\bm{O}$ at the end.  The formalism developed in
this note is completely transparent with respect to this
factorization.  In particular:
\begin{itemize}
\item the full Hamiltonian $\mathring{\bm H}(E)$ can be decomposed as
  $\mathring{\bm H}(\bar{E} + \xi_E) \approx \ringbar{\bm H} +
  \mathring{\bm H}_E'\, \xi_E$ with $\ringbar{\bm H} \equiv \bm{O}\,
  \bar{\bm H}\, \bm{O}^\dagger$ and $\mathring{\bm H}_E' \equiv
  \bm{O}\, \bm{H}_E'\, \bm{O}^\dagger$.  Similarly, $\ringbar{\bm S}
  \equiv \bm{O}\, \bar{\bm S}\, \bm{O}^\dagger$;

\item the Hamiltonian $\ringbar{\bm H}$ is diagonalized by a matrix
  $\ringbar{\bm U} \equiv \bm{O} \, \bar{\bm U}$, and its eigenvalues
  are the same as $\bar{\bm H}$.  Hence, $\ringbar{\bm U}^\dagger\,
  \ringbar{\bm H}\, \ringbar{\bm U} = \bar{\bm U}^\dagger\, \bar{\bm
    H}\, \bar{\bm U} = \bar{\bm\omega}$.  Consequently, the matrix
  $\bm{C}$ used to construct $\bm{K}_E$ is unaffected by $\bm{O}$, and
  we get $\mathring{\bm K}_E = \bm{O}\, \bm{K}_E\, \bm{O}^\dagger$ as
  expected;

\item in brief, the pair $(\ringbar{\bm S}, \mathring{\bm K}_E)$
  accounting for the full Hamiltonian $\mathring{\bm H}(E)$ is related
  to the reduced one $(\bar{\bm S}, \bm{K}_E)$ by an overall rotation
  of each individual matrix: $(\ringbar{\bm S}, \mathring{\bm K}_E)
  \equiv \bm{O}\, (\bar{\bm S}, \bm{K}_E)\, \bm{O}^\dagger = (\bm{O}\,
  \bar{\bm S}\, \bm{O}^\dagger, \bm{O}\, \bm{K}_E\, \bm{O}^\dagger)$.
  The same happens when multiple derivatives (such as $\bm{K}_E$ and
  $\bm{K}_\Theta$) or the altitude of the production point for
  atmospheric neutrinos are considered.
\end{itemize}
As for the actual averaging, the algorithm described in
section~\ref{sec:altitude} is trivially modified in its first and last
step to incorporate $\bm{O}$:
\begin{enumerate}[label=\emph{\alph*})]
\item the initial matrix $\bm\rho_0$ must be rotated to the
  propagation basis, so that:
  \begin{equation}
    \bm\rho_0 \equiv \bm{O}^\dagger\, \bm\Pi^{(\alpha)}\, \bm{O}
    \quad\text{with}\quad
    \bm\Pi^{(\alpha)}_{ij} = \delta_{\alpha i} \delta_{\alpha j} \,;
  \end{equation}

\item[\emph{b--d})] remain the same as before, expressed in terms of
  the matrices ``without the $\bm{O}$'';

\item[\emph{e})] the matrix $\bm{O}$ is reintroduced in the
  construction of the final density matrix $\bm\rho$:
  \begin{equation}
    \bm\rho_3 \to \bm\rho \equiv \bm{O}\,
    \bar{\bm S}\, \bm\rho_3\, \bar{\bm S}^\dagger\,
    \bm{O}^\dagger.
  \end{equation}
\end{enumerate}
In summary, models where a group of parameters can be factorized out
and reintroduced at the end are perfectly compatible with our
formalism.

Finally, we want to comment on a rather common feature of many
oscillation models which is sometimes exploited to speed-up
computations.  It is not infrequent that the oscillation probabilities
depend on the neutrino energy only through a particular combination of
it with the parameters of the model (here collectively denoted as
$\vec\Omega$), so that $P$ is invariant under a simultaneous rescaling
of the energy $E\to \alpha E$ and suitable transformation of the
parameters $\vec\Omega \to \vec\Omega_\alpha$.  This is the case, for
example, in standard three-neutrino oscillations, where the
probabilities depend on the energy and the mass-squared differences
through the combined ratios $\Dmq_{ij} / E$.  Such situation allows to
reuse the probability spectrum $P(E)$ tabulated for a given point
$\vec\Omega$ in parameter space, for all the other points related to
it by the transformation $\vec\Omega \to \vec\Omega_\alpha$, as these
would require a simple ``shift'' $E \to \alpha E$ of the tabulated
energy values.  When energy averaging is introduced into the game, one
should be careful not to spoil the invariance of the system.
Concretely, denoting by $\Delta[\bar{E}]$ the range of the bin with
central energy $\bar{E}$, we should make sure that
$\Delta[\alpha\bar{E}] = \alpha\, \Delta[\bar{E}]$.  This is trivially
achieved if $\Delta[\bar{E}] \mathbin{\big/} \bar{E} =
\text{constant}$, \textit{i.e.}, a uniform spacing in $\log(E)$ is
used to define the energy grid.

\bibliographystyle{JHEPmod}
\bibliography{references}

\providecommand{\href}[2]{#2}\begingroup\raggedright\begin{thebibliography}{100}

\bibitem{Honda:2015fha}
M.~Honda, M.~Sajjad~Athar, T.~Kajita, K.~Kasahara and S.~Midorikawa,
  \emph{{Atmospheric neutrino flux calculation using the NRLMSISE-00
  atmospheric model}},
  \href{https://doi.org/10.1103/PhysRevD.92.023004}{\emph{Phys. Rev.}
  {\bfseries D92} (2015) 023004}
  [\href{https://arxiv.org/abs/1502.03916}{{\ttfamily 1502.03916}}].

\bibitem{Vinyoles:2016djt}
N.~Vinyoles, A.M.~Serenelli, F.L.~Villante, S.~Basu, J.~Bergström,
  M.C.~Gonzalez-Garcia et~al., \emph{{A new Generation of Standard Solar
  Models}}, \href{https://doi.org/10.3847/1538-4357/835/2/202}{\emph{Astrophys.
  J.} {\bfseries 835} (2017) 202}
  [\href{https://arxiv.org/abs/1611.09867}{{\ttfamily 1611.09867}}].

\bibitem{Arafune:1996bt}
J.~Arafune and J.~Sato, \emph{{CP and T violation test in neutrino
  oscillation}}, \href{https://doi.org/10.1103/PhysRevD.55.1653}{\emph{Phys.
  Rev. D} {\bfseries 55} (1997) 1653}
  [\href{https://arxiv.org/abs/hep-ph/9607437}{{\ttfamily hep-ph/9607437}}].

\bibitem{Lisi:1997yc}
E.~Lisi and D.~Montanino, \emph{{Earth regeneration effect in solar neutrino
  oscillations: An Analytic approach}},
  \href{https://doi.org/10.1103/PhysRevD.56.1792}{\emph{Phys. Rev. D}
  {\bfseries 56} (1997) 1792}
  [\href{https://arxiv.org/abs/hep-ph/9702343}{{\ttfamily hep-ph/9702343}}].

\bibitem{Arafune:1997hd}
J.~Arafune, M.~Koike and J.~Sato, \emph{{CP violation and matter effect in long
  baseline neutrino oscillation experiments}},
  \href{https://doi.org/10.1103/PhysRevD.60.119905}{\emph{Phys. Rev. D}
  {\bfseries 56} (1997) 3093}
  [\href{https://arxiv.org/abs/hep-ph/9703351}{{\ttfamily hep-ph/9703351}}].
  [Erratum: Phys.Rev.D 60, 119905 (1999)].

\bibitem{Minakata:1998bf}
H.~Minakata and H.~Nunokawa, \emph{{CP violation versus matter effect in long
  baseline neutrino oscillation experiments}},
  \href{https://doi.org/10.1103/PhysRevD.57.4403}{\emph{Phys. Rev. D}
  {\bfseries 57} (1998) 4403}
  [\href{https://arxiv.org/abs/hep-ph/9705208}{{\ttfamily hep-ph/9705208}}].

\bibitem{Petcov:1998su}
S.T.~Petcov, \emph{{Diffractive - like (or parametric resonance - like?)
  enhancement of the earth (day - night) effect for solar neutrinos crossing
  the earth core}},
  \href{https://doi.org/10.1016/S0370-2693(98)00742-4}{\emph{Phys. Lett. B}
  {\bfseries 434} (1998) 321}
  [\href{https://arxiv.org/abs/hep-ph/9805262}{{\ttfamily hep-ph/9805262}}].

\bibitem{Akhmedov:1998ui}
E.K.~Akhmedov, \emph{{Parametric resonance of neutrino oscillations and passage
  of solar and atmospheric neutrinos through the earth}},
  \href{https://doi.org/10.1016/S0550-3213(98)00723-8}{\emph{Nucl. Phys. B}
  {\bfseries 538} (1999) 25}
  [\href{https://arxiv.org/abs/hep-ph/9805272}{{\ttfamily hep-ph/9805272}}].

\bibitem{Akhmedov:1998xq}
E.K.~Akhmedov, A.~Dighe, P.~Lipari and A.Y.~Smirnov, \emph{{Atmospheric
  neutrinos at Super-Kamiokande and parametric resonance in neutrino
  oscillations}},
  \href{https://doi.org/10.1016/S0550-3213(98)00825-6}{\emph{Nucl. Phys. B}
  {\bfseries 542} (1999) 3}
  [\href{https://arxiv.org/abs/hep-ph/9808270}{{\ttfamily hep-ph/9808270}}].

\bibitem{Chizhov:1999he}
M.V.~Chizhov and S.T.~Petcov, \emph{{Enhancing mechanisms of neutrino
  transitions in a medium of nonperiodic constant density layers and in the
  earth}}, \href{https://doi.org/10.1103/PhysRevD.63.073003}{\emph{Phys. Rev.
  D} {\bfseries 63} (2001) 073003}
  [\href{https://arxiv.org/abs/hep-ph/9903424}{{\ttfamily hep-ph/9903424}}].

\bibitem{Ohlsson:1999xb}
T.~Ohlsson and H.~Snellman, \emph{{Three flavor neutrino oscillations in
  matter}}, \href{https://doi.org/10.1063/1.533270}{\emph{J. Math. Phys.}
  {\bfseries 41} (2000) 2768}
  [\href{https://arxiv.org/abs/hep-ph/9910546}{{\ttfamily hep-ph/9910546}}].
  [Erratum: J.Math.Phys. 42, 2345 (2001)].

\bibitem{Koike:1999tb}
M.~Koike and J.~Sato, \emph{{T violation search with very long baseline
  neutrino oscillation experiments}},
  \href{https://doi.org/10.1103/PhysRevD.62.073006}{\emph{Phys. Rev. D}
  {\bfseries 62} (2000) 073006}
  [\href{https://arxiv.org/abs/hep-ph/9911258}{{\ttfamily hep-ph/9911258}}].

\bibitem{Ohlsson:1999um}
T.~Ohlsson and H.~Snellman, \emph{{Neutrino oscillations with three flavors in
  matter: Applications to neutrinos traversing the Earth}},
  \href{https://doi.org/10.1016/S0370-2693(00)00008-3}{\emph{Phys. Lett. B}
  {\bfseries 474} (2000) 153}
  [\href{https://arxiv.org/abs/hep-ph/9912295}{{\ttfamily hep-ph/9912295}}].
  [Erratum: Phys.Lett.B 480, 419--419 (2000)].

\bibitem{Harrison:1999df}
P.F.~Harrison and W.G.~Scott, \emph{{CP and T violation in neutrino
  oscillations and invariance of Jarlskog's determinant to matter effects}},
  \href{https://doi.org/10.1016/S0370-2693(00)00153-2}{\emph{Phys. Lett. B}
  {\bfseries 476} (2000) 349}
  [\href{https://arxiv.org/abs/hep-ph/9912435}{{\ttfamily hep-ph/9912435}}].

\bibitem{DeRujula:2000ap}
A.~De~Rujula, M.B.~Gavela and P.~Hernandez, \emph{{The atmospheric neutrino
  anomaly without maximal mixing?}},
  \href{https://doi.org/10.1103/PhysRevD.63.033001}{\emph{Phys. Rev. D}
  {\bfseries 63} (2001) 033001}
  [\href{https://arxiv.org/abs/hep-ph/0001124}{{\ttfamily hep-ph/0001124}}].

\bibitem{Cervera:2000kp}
A.~Cervera, A.~Donini, M.B.~Gavela, J.J.~Gomez~Cadenas, P.~Hernandez, O.~Mena
  et~al., \emph{{Golden measurements at a neutrino factory}},
  \href{https://doi.org/10.1016/S0550-3213(00)00221-2}{\emph{Nucl. Phys. B}
  {\bfseries 579} (2000) 17}
  [\href{https://arxiv.org/abs/hep-ph/0002108}{{\ttfamily hep-ph/0002108}}].
  [Erratum: Nucl.Phys.B 593, 731--732 (2001)].

\bibitem{Lunardini:2000swa}
C.~Lunardini and A.Y.~Smirnov, \emph{{The Minimum width condition for neutrino
  conversion in matter}},
  \href{https://doi.org/10.1016/S0550-3213(00)00341-2}{\emph{Nucl. Phys. B}
  {\bfseries 583} (2000) 260}
  [\href{https://arxiv.org/abs/hep-ph/0002152}{{\ttfamily hep-ph/0002152}}].

\bibitem{Minakata:2000ee}
H.~Minakata and H.~Nunokawa, \emph{{Measuring leptonic CP violation by
  low-energy neutrino oscillation experiments}},
  \href{https://doi.org/10.1016/S0370-2693(00)01249-1}{\emph{Phys. Lett. B}
  {\bfseries 495} (2000) 369}
  [\href{https://arxiv.org/abs/hep-ph/0004114}{{\ttfamily hep-ph/0004114}}].

\bibitem{Akhmedov:2000cs}
E.K.~Akhmedov, \emph{{Matter effects in oscillations of neutrinos traveling
  short distances in matter}},
  \href{https://doi.org/10.1016/S0370-2693(01)00165-4}{\emph{Phys. Lett. B}
  {\bfseries 503} (2001) 133}
  [\href{https://arxiv.org/abs/hep-ph/0011136}{{\ttfamily hep-ph/0011136}}].

\bibitem{Lisi:2000su}
E.~Lisi, A.~Marrone, D.~Montanino, A.~Palazzo and S.T.~Petcov,
  \emph{{Analytical description of quasivacuum oscillations of solar
  neutrinos}}, \href{https://doi.org/10.1103/PhysRevD.63.093002}{\emph{Phys.
  Rev. D} {\bfseries 63} (2001) 093002}
  [\href{https://arxiv.org/abs/hep-ph/0011306}{{\ttfamily hep-ph/0011306}}].

\bibitem{Ohlsson:2001et}
T.~Ohlsson and H.~Snellman, \emph{{Neutrino oscillations with three flavors in
  matter of varying density}},
  \href{https://doi.org/10.1007/s100520100687}{\emph{Eur. Phys. J. C}
  {\bfseries 20} (2001) 507}
  [\href{https://arxiv.org/abs/hep-ph/0103252}{{\ttfamily hep-ph/0103252}}].

\bibitem{Freund:2001pn}
M.~Freund, \emph{{Analytic approximations for three neutrino oscillation
  parameters and probabilities in matter}},
  \href{https://doi.org/10.1103/PhysRevD.64.053003}{\emph{Phys. Rev. D}
  {\bfseries 64} (2001) 053003}
  [\href{https://arxiv.org/abs/hep-ph/0103300}{{\ttfamily hep-ph/0103300}}].

\bibitem{Akhmedov:2001kd}
E.K.~Akhmedov, P.~Huber, M.~Lindner and T.~Ohlsson, \emph{{T violation in
  neutrino oscillations in matter}},
  \href{https://doi.org/10.1016/S0550-3213(01)00261-9}{\emph{Nucl. Phys. B}
  {\bfseries 608} (2001) 394}
  [\href{https://arxiv.org/abs/hep-ph/0105029}{{\ttfamily hep-ph/0105029}}].

\bibitem{Yasuda:2001va}
O.~Yasuda, \emph{{Vacuum mimicking phenomena in neutrino oscillations}},
  \href{https://doi.org/10.1016/S0370-2693(01)00920-0}{\emph{Phys. Lett. B}
  {\bfseries 516} (2001) 111}
  [\href{https://arxiv.org/abs/hep-ph/0106232}{{\ttfamily hep-ph/0106232}}].

\bibitem{Kimura:2002hb}
K.~Kimura, A.~Takamura and H.~Yokomakura, \emph{{Exact formula of probability
  and CP violation for neutrino oscillations in matter}},
  \href{https://doi.org/10.1016/S0370-2693(02)01907-X}{\emph{Phys. Lett. B}
  {\bfseries 537} (2002) 86}
  [\href{https://arxiv.org/abs/hep-ph/0203099}{{\ttfamily hep-ph/0203099}}].

\bibitem{Kimura:2002wd}
K.~Kimura, A.~Takamura and H.~Yokomakura, \emph{{Exact formulas and simple CP
  dependence of neutrino oscillation probabilities in matter with constant
  density}}, \href{https://doi.org/10.1103/PhysRevD.66.073005}{\emph{Phys. Rev.
  D} {\bfseries 66} (2002) 073005}
  [\href{https://arxiv.org/abs/hep-ph/0205295}{{\ttfamily hep-ph/0205295}}].

\bibitem{Yokomakura:2002av}
H.~Yokomakura, K.~Kimura and A.~Takamura, \emph{{Overall feature of CP
  dependence for neutrino oscillation probability in arbitrary matter
  profile}}, \href{https://doi.org/10.1016/S0370-2693(02)02545-5}{\emph{Phys.
  Lett. B} {\bfseries 544} (2002) 286}
  [\href{https://arxiv.org/abs/hep-ph/0207174}{{\ttfamily hep-ph/0207174}}].

\bibitem{Aguilar-Arevalo:2003hty}
A.A.~Aguilar-Arevalo, L.G.~Cabral-Rosetti and J.C.~D'Olivo, \emph{{Magnus
  expansion and three neutrino oscillations in matter}},
  \href{https://doi.org/10.1088/1742-6596/37/1/028}{\emph{J. Phys. Conf. Ser.}
  {\bfseries 37} (2006) 161}
  [\href{https://arxiv.org/abs/hep-ph/0302017}{{\ttfamily hep-ph/0302017}}].

\bibitem{Jacobson:2003wc}
M.~Jacobson and T.~Ohlsson, \emph{{Extrinsic CPT violation in neutrino
  oscillations in matter}},
  \href{https://doi.org/10.1103/PhysRevD.69.013003}{\emph{Phys. Rev. D}
  {\bfseries 69} (2004) 013003}
  [\href{https://arxiv.org/abs/hep-ph/0305064}{{\ttfamily hep-ph/0305064}}].

\bibitem{Akhmedov:2004ny}
E.K.~Akhmedov, R.~Johansson, M.~Lindner, T.~Ohlsson and T.~Schwetz,
  \emph{{Series expansions for three flavor neutrino oscillation probabilities
  in matter}}, \href{https://doi.org/10.1088/1126-6708/2004/04/078}{\emph{JHEP}
  {\bfseries 04} (2004) 078}
  [\href{https://arxiv.org/abs/hep-ph/0402175}{{\ttfamily hep-ph/0402175}}].

\bibitem{deHolanda:2004fd}
P.C.~de~Holanda, W.~Liao and A.Y.~Smirnov, \emph{{Toward precision measurements
  in solar neutrinos}},
  \href{https://doi.org/10.1016/j.nuclphysb.2004.09.027}{\emph{Nucl. Phys. B}
  {\bfseries 702} (2004) 307}
  [\href{https://arxiv.org/abs/hep-ph/0404042}{{\ttfamily hep-ph/0404042}}].

\bibitem{Ioannisian:2004jk}
A.N.~Ioannisian and A.Y.~Smirnov, \emph{{Neutrino oscillations in low density
  medium}}, \href{https://doi.org/10.1103/PhysRevLett.93.241801}{\emph{Phys.
  Rev. Lett.} {\bfseries 93} (2004) 241801}
  [\href{https://arxiv.org/abs/hep-ph/0404060}{{\ttfamily hep-ph/0404060}}].

\bibitem{Akhmedov:2004rq}
E.K.~Akhmedov, M.A.~Tortola and J.W.F.~Valle, \emph{{A Simple analytic three
  flavor description of the day night effect in the solar neutrino flux}},
  \href{https://doi.org/10.1088/1126-6708/2004/05/057}{\emph{JHEP} {\bfseries
  05} (2004) 057} [\href{https://arxiv.org/abs/hep-ph/0404083}{{\ttfamily
  hep-ph/0404083}}].

\bibitem{Blennow:2004qd}
M.~Blennow and T.~Ohlsson, \emph{{Exact series solution to the two flavor
  neutrino oscillation problem in matter}},
  \href{https://doi.org/10.1063/1.1793330}{\emph{J. Math. Phys.} {\bfseries 45}
  (2004) 4053} [\href{https://arxiv.org/abs/hep-ph/0405033}{{\ttfamily
  hep-ph/0405033}}].

\bibitem{Ioannisian:2004vv}
A.N.~Ioannisian, N.A.~Kazarian, A.Y.~Smirnov and D.~Wyler, \emph{{A Precise
  analytical description of the earth matter effect on oscillations of low
  energy neutrinos}},
  \href{https://doi.org/10.1103/PhysRevD.71.033006}{\emph{Phys. Rev. D}
  {\bfseries 71} (2005) 033006}
  [\href{https://arxiv.org/abs/hep-ph/0407138}{{\ttfamily hep-ph/0407138}}].

\bibitem{Akhmedov:2005yj}
E.K.~Akhmedov, M.~Maltoni and A.Y.~Smirnov, \emph{{Oscillations of high energy
  neutrinos in matter: Precise formalism and parametric resonance}},
  \href{https://doi.org/10.1103/PhysRevLett.95.211801}{\emph{Phys. Rev. Lett.}
  {\bfseries 95} (2005) 211801}
  [\href{https://arxiv.org/abs/hep-ph/0506064}{{\ttfamily hep-ph/0506064}}].

\bibitem{Takamura:2005df}
A.~Takamura and K.~Kimura, \emph{{Large non-perturbative effects of small
  $\Delta m^2_{21} / \Delta m^2_{31}$ and $\sin\theta_{13}$ on neutrino
  oscillation and CP violation in matter}},
  \href{https://doi.org/10.1088/1126-6708/2006/01/053}{\emph{JHEP} {\bfseries
  01} (2006) 053} [\href{https://arxiv.org/abs/hep-ph/0506112}{{\ttfamily
  hep-ph/0506112}}].

\bibitem{Choubey:2005zy}
S.~Choubey and P.~Roy, \emph{{Probing the deviation from maximal mixing of
  atmospheric neutrinos}},
  \href{https://doi.org/10.1103/PhysRevD.73.013006}{\emph{Phys. Rev. D}
  {\bfseries 73} (2006) 013006}
  [\href{https://arxiv.org/abs/hep-ph/0509197}{{\ttfamily hep-ph/0509197}}].

\bibitem{Akhmedov:2006hb}
E.K.~Akhmedov, M.~Maltoni and A.Y.~Smirnov, \emph{{1-3 leptonic mixing and the
  neutrino oscillograms of the Earth}},
  \href{https://doi.org/10.1088/1126-6708/2007/05/077}{\emph{JHEP} {\bfseries
  05} (2007) 077} [\href{https://arxiv.org/abs/hep-ph/0612285}{{\ttfamily
  hep-ph/0612285}}].

\bibitem{deAquino:2007sx}
V.M.~de~Aquino and J.S.S.~de~Oliveira, \emph{{Aproximative solutions to the
  neutrino oscillation problem in matter}},
  \href{https://doi.org/10.1088/0031-8949/77/04/045101}{\emph{Phys. Scripta}
  {\bfseries 77} (2008) 045101}
  [\href{https://arxiv.org/abs/hep-ph/0703151}{{\ttfamily hep-ph/0703151}}].

\bibitem{Liao:2007re}
W.~Liao, \emph{{Precise Formulation of Neutrino Oscillation in the Earth}},
  \href{https://doi.org/10.1103/PhysRevD.77.053002}{\emph{Phys. Rev. D}
  {\bfseries 77} (2008) 053002}
  [\href{https://arxiv.org/abs/0710.1492}{{\ttfamily 0710.1492}}].

\bibitem{Ioannisian:2008ve}
A.N.~Ioannisian and A.Y.~Smirnov, \emph{{Describing neutrino oscillations in
  matter with Magnus expansion}},
  \href{https://doi.org/10.1016/j.nuclphysb.2009.02.028}{\emph{Nucl. Phys. B}
  {\bfseries 816} (2009) 94} [\href{https://arxiv.org/abs/0803.1967}{{\ttfamily
  0803.1967}}].

\bibitem{Akhmedov:2008qt}
E.K.~Akhmedov, M.~Maltoni and A.Y.~Smirnov, \emph{{Neutrino oscillograms of the
  Earth: Effects of 1-2 mixing and CP-violation}},
  \href{https://doi.org/10.1088/1126-6708/2008/06/072}{\emph{JHEP} {\bfseries
  06} (2008) 072} [\href{https://arxiv.org/abs/0804.1466}{{\ttfamily
  0804.1466}}].

\bibitem{Kikuchi:2008vq}
T.~Kikuchi, H.~Minakata and S.~Uchinami, \emph{{Perturbation Theory of Neutrino
  Oscillation with Nonstandard Neutrino Interactions}},
  \href{https://doi.org/10.1088/1126-6708/2009/03/114}{\emph{JHEP} {\bfseries
  03} (2009) 114} [\href{https://arxiv.org/abs/0809.3312}{{\ttfamily
  0809.3312}}].

\bibitem{Asano:2011nj}
K.~Asano and H.~Minakata, \emph{{Large-$\theta_{13}$ Perturbation Theory of
  Neutrino Oscillation for Long-Baseline Experiments}},
  \href{https://doi.org/10.1007/JHEP06(2011)022}{\emph{JHEP} {\bfseries 06}
  (2011) 022} [\href{https://arxiv.org/abs/1103.4387}{{\ttfamily 1103.4387}}].

\bibitem{Agarwalla:2013tza}
S.K.~Agarwalla, Y.~Kao and T.~Takeuchi, \emph{{Analytical approximation of the
  neutrino oscillation matter effects at large $\theta_{13}$}},
  \href{https://doi.org/10.1007/JHEP04(2014)047}{\emph{JHEP} {\bfseries 04}
  (2014) 047} [\href{https://arxiv.org/abs/1302.6773}{{\ttfamily 1302.6773}}].

\bibitem{Blennow:2013rca}
M.~Blennow and A.Y.~Smirnov, \emph{{Neutrino propagation in matter}},
  \href{https://doi.org/10.1155/2013/972485}{\emph{Adv. High Energy Phys.}
  {\bfseries 2013} (2013) 972485}
  [\href{https://arxiv.org/abs/1306.2903}{{\ttfamily 1306.2903}}].

\bibitem{Coloma:2014kca}
P.~Coloma, H.~Minakata and S.J.~Parke, \emph{{Interplay between appearance and
  disappearance channels for precision measurements of $\theta_{23}$ and
  $\delta$}}, \href{https://doi.org/10.1103/PhysRevD.90.093003}{\emph{Phys.
  Rev. D} {\bfseries 90} (2014) 093003}
  [\href{https://arxiv.org/abs/1406.2551}{{\ttfamily 1406.2551}}].

\bibitem{Xu:2015kma}
X.-J.~Xu, \emph{{Why is the neutrino oscillation formula expanded in
  \ensuremath{\Delta}m$_{21}^{2}$ /\ensuremath{\Delta}m$_{31}^{2}$ still
  accurate near the solar resonance in matter?}},
  \href{https://doi.org/10.1007/JHEP10(2015)090}{\emph{JHEP} {\bfseries 10}
  (2015) 090} [\href{https://arxiv.org/abs/1502.02503}{{\ttfamily
  1502.02503}}].

\bibitem{Minakata:2015gra}
H.~Minakata and S.J.~Parke, \emph{{Simple and Compact Expressions for Neutrino
  Oscillation Probabilities in Matter}},
  \href{https://doi.org/10.1007/JHEP01(2016)180}{\emph{JHEP} {\bfseries 01}
  (2016) 180} [\href{https://arxiv.org/abs/1505.01826}{{\ttfamily
  1505.01826}}].

\bibitem{Parke:2016joa}
S.~Parke, \emph{{What is $\Delta m^2_{ee}$?}},
  \href{https://doi.org/10.1103/PhysRevD.93.053008}{\emph{Phys. Rev. D}
  {\bfseries 93} (2016) 053008}
  [\href{https://arxiv.org/abs/1601.07464}{{\ttfamily 1601.07464}}].

\bibitem{Denton:2016wmg}
P.B.~Denton, H.~Minakata and S.J.~Parke, \emph{{Compact Perturbative
  Expressions For Neutrino Oscillations in Matter}},
  \href{https://doi.org/10.1007/JHEP06(2016)051}{\emph{JHEP} {\bfseries 06}
  (2016) 051} [\href{https://arxiv.org/abs/1604.08167}{{\ttfamily
  1604.08167}}].

\bibitem{Ge:2016dlx}
S.-F.~Ge and A.Y.~Smirnov, \emph{{Non-standard interactions and the CP phase
  measurements in neutrino oscillations at low energies}},
  \href{https://doi.org/10.1007/JHEP10(2016)138}{\emph{JHEP} {\bfseries 10}
  (2016) 138} [\href{https://arxiv.org/abs/1607.08513}{{\ttfamily
  1607.08513}}].

\bibitem{Fong:2016yyh}
C.S.~Fong, H.~Minakata and H.~Nunokawa, \emph{{A framework for testing leptonic
  unitarity by neutrino oscillation experiments}},
  \href{https://doi.org/10.1007/JHEP02(2017)114}{\emph{JHEP} {\bfseries 02}
  (2017) 114} [\href{https://arxiv.org/abs/1609.08623}{{\ttfamily
  1609.08623}}].

\bibitem{Li:2016pzm}
Y.-F.~Li, J.~Zhang, S.~Zhou and J.-y.~Zhu, \emph{{Looking into Analytical
  Approximations for Three-flavor Neutrino Oscillation Probabilities in
  Matter}}, \href{https://doi.org/10.1007/JHEP12(2016)109}{\emph{JHEP}
  {\bfseries 12} (2016) 109}
  [\href{https://arxiv.org/abs/1610.04133}{{\ttfamily 1610.04133}}].

\bibitem{Pas:2016qbg}
H.~P\"as and P.~Sicking, \emph{{Discriminating sterile neutrinos and unitarity
  violation with CP invariants}},
  \href{https://doi.org/10.1103/PhysRevD.95.075004}{\emph{Phys. Rev. D}
  {\bfseries 95} (2017) 075004}
  [\href{https://arxiv.org/abs/1611.08450}{{\ttfamily 1611.08450}}].

\bibitem{Minakata:2017ahk}
H.~Minakata, \emph{{An Effective Two-Flavor Approximation for Neutrino Survival
  Probabilities in Matter}},
  \href{https://doi.org/10.1007/JHEP05(2017)043}{\emph{JHEP} {\bfseries 05}
  (2017) 043} [\href{https://arxiv.org/abs/1702.03332}{{\ttfamily
  1702.03332}}].

\bibitem{Fong:2017gke}
C.S.~Fong, H.~Minakata and H.~Nunokawa, \emph{{Non-unitary evolution of
  neutrinos in matter and the leptonic unitarity test}},
  \href{https://doi.org/10.1007/JHEP02(2019)015}{\emph{JHEP} {\bfseries 02}
  (2019) 015} [\href{https://arxiv.org/abs/1712.02798}{{\ttfamily
  1712.02798}}].

\bibitem{Denton:2018hal}
P.B.~Denton and S.J.~Parke, \emph{{Addendum to ``Compact perturbative
  expressions for neutrino oscillations in matter''}},
  \href{https://arxiv.org/abs/1801.06514}{{\ttfamily 1801.06514}}. [Addendum:
  JHEP 06, 109 (2018)].

\bibitem{Ioannisian:2018qwl}
A.~Ioannisian and S.~Pokorski, \emph{{Three Neutrino Oscillations in Matter}},
  \href{https://doi.org/10.1016/j.physletb.2018.06.001}{\emph{Phys. Lett. B}
  {\bfseries 782} (2018) 641}
  [\href{https://arxiv.org/abs/1801.10488}{{\ttfamily 1801.10488}}].

\bibitem{Denton:2018fex}
P.B.~Denton, S.J.~Parke and X.~Zhang, \emph{{Rotations Versus Perturbative
  Expansions for Calculating Neutrino Oscillation Probabilities in Matter}},
  \href{https://doi.org/10.1103/PhysRevD.98.033001}{\emph{Phys. Rev. D}
  {\bfseries 98} (2018) 033001}
  [\href{https://arxiv.org/abs/1806.01277}{{\ttfamily 1806.01277}}].

\bibitem{Bernabeu:2018twl}
J.~Bernab\'eu and A.~Segarra, \emph{{Disentangling genuine from matter-induced
  CP violation in neutrino oscillations}},
  \href{https://doi.org/10.1103/PhysRevLett.121.211802}{\emph{Phys. Rev. Lett.}
  {\bfseries 121} (2018) 211802}
  [\href{https://arxiv.org/abs/1806.07694}{{\ttfamily 1806.07694}}].

\bibitem{Martinez-Soler:2018lcy}
I.~Martinez-Soler and H.~Minakata, \emph{{Standard versus Non-Standard CP
  Phases in Neutrino Oscillation in Matter with Non-Unitarity}},
  \href{https://doi.org/10.1093/ptep/ptaa062}{\emph{PTEP} {\bfseries 2020}
  (2020) 063B01} [\href{https://arxiv.org/abs/1806.10152}{{\ttfamily
  1806.10152}}].

\bibitem{Bernabeu:2018use}
J.~Bernab\'eu and A.~Segarra, \emph{{Signatures of the genuine and
  matter-induced components of the CP violation asymmetry in neutrino
  oscillations}}, \href{https://doi.org/10.1007/JHEP11(2018)063}{\emph{JHEP}
  {\bfseries 11} (2018) 063}
  [\href{https://arxiv.org/abs/1807.11879}{{\ttfamily 1807.11879}}].

\bibitem{Li:2018ezt}
W.~Li, J.~Ling, F.~Xu and B.~Yue, \emph{{Matter Effect of Light Sterile
  Neutrino: An Exact Analytical Approach}},
  \href{https://doi.org/10.1007/JHEP10(2018)021}{\emph{JHEP} {\bfseries 10}
  (2018) 021} [\href{https://arxiv.org/abs/1808.03985}{{\ttfamily
  1808.03985}}].

\bibitem{Denton:2018cpu}
P.B.~Denton and S.J.~Parke, \emph{{The effective $\Delta m^2_{ee}$ in matter}},
  \href{https://doi.org/10.1103/PhysRevD.98.093001}{\emph{Phys. Rev. D}
  {\bfseries 98} (2018) 093001}
  [\href{https://arxiv.org/abs/1808.09453}{{\ttfamily 1808.09453}}].

\bibitem{Chaves:2018sih}
M.E.~Chaves, D.R.~Gratieri and O.L.G.~Peres, \emph{{Improvements on
  perturbative oscillation formulas including non-standard neutrino
  interactions}}, \href{https://doi.org/10.1088/1361-6471/abae17}{\emph{J.
  Phys. G} {\bfseries 48} (2020) 015001}
  [\href{https://arxiv.org/abs/1810.04979}{{\ttfamily 1810.04979}}].

\bibitem{Barenboim:2019pfp}
G.~Barenboim, P.B.~Denton, S.J.~Parke and C.A.~Ternes, \emph{{Neutrino
  Oscillation Probabilities through the Looking Glass}},
  \href{https://doi.org/10.1016/j.physletb.2019.03.002}{\emph{Phys. Lett. B}
  {\bfseries 791} (2019) 351}
  [\href{https://arxiv.org/abs/1902.00517}{{\ttfamily 1902.00517}}].

\bibitem{Denton:2019yiw}
P.B.~Denton and S.J.~Parke, \emph{{Simple and Precise Factorization of the
  Jarlskog Invariant for Neutrino Oscillations in Matter}},
  \href{https://doi.org/10.1103/PhysRevD.100.053004}{\emph{Phys. Rev. D}
  {\bfseries 100} (2019) 053004}
  [\href{https://arxiv.org/abs/1902.07185}{{\ttfamily 1902.07185}}].

\bibitem{Martinez-Soler:2019nhb}
I.~Martinez-Soler and H.~Minakata, \emph{{Perturbing Neutrino Oscillations
  Around the Solar Resonance}},
  \href{https://doi.org/10.1093/ptep/ptz067}{\emph{PTEP} {\bfseries 2019}
  (2019) 073B07} [\href{https://arxiv.org/abs/1904.07853}{{\ttfamily
  1904.07853}}].

\bibitem{Parke:2019jyu}
S.J.~Parke and X.~Zhang, \emph{{Compact Perturbative Expressions for
  Oscillations with Sterile Neutrinos in Matter}},
  \href{https://doi.org/10.1103/PhysRevD.101.056005}{\emph{Phys. Rev. D}
  {\bfseries 101} (2020) 056005}
  [\href{https://arxiv.org/abs/1905.01356}{{\ttfamily 1905.01356}}].

\bibitem{Denton:2019ovn}
P.B.~Denton, S.J.~Parke and X.~Zhang, \emph{{Neutrino oscillations in matter
  via eigenvalues}},
  \href{https://doi.org/10.1103/PhysRevD.101.093001}{\emph{Phys. Rev. D}
  {\bfseries 101} (2020) 093001}
  [\href{https://arxiv.org/abs/1907.02534}{{\ttfamily 1907.02534}}].

\bibitem{Martinez-Soler:2019noy}
I.~Martinez-Soler and H.~Minakata, \emph{{Physics of parameter correlations
  around the solar-scale enhancement in neutrino theory with unitarity
  violation}}, \href{https://doi.org/10.1093/ptep/ptaa112}{\emph{PTEP}
  {\bfseries 2020} (2020) 113B01}
  [\href{https://arxiv.org/abs/1908.04855}{{\ttfamily 1908.04855}}].

\bibitem{Wang:2019dal}
X.~Wang and S.~Zhou, \emph{{On the Properties of the Effective Jarlskog
  Invariant for Three-flavor Neutrino Oscillations in Matter}},
  \href{https://doi.org/10.1016/j.nuclphysb.2019.114867}{\emph{Nucl. Phys. B}
  {\bfseries 950} (2020) 114867}
  [\href{https://arxiv.org/abs/1908.07304}{{\ttfamily 1908.07304}}].

\bibitem{Denton:2019qzn}
P.B.~Denton, S.J.~Parke and X.~Zhang, \emph{{Fibonacci Fast Convergence for
  Neutrino Oscillations in Matter}},
  \href{https://doi.org/10.1016/j.physletb.2020.135592}{\emph{Phys. Lett. B}
  {\bfseries 807} (2020) 135592}
  [\href{https://arxiv.org/abs/1909.02009}{{\ttfamily 1909.02009}}].

\bibitem{Luo:2019efb}
S.~Luo, \emph{{Neutrino Oscillation in Dense Matter}},
  \href{https://doi.org/10.1103/PhysRevD.101.033005}{\emph{Phys. Rev. D}
  {\bfseries 101} (2020) 033005}
  [\href{https://arxiv.org/abs/1911.06301}{{\ttfamily 1911.06301}}].

\bibitem{Huber:2019frh}
P.~Huber, H.~Minakata and R.~Pestes, \emph{{Interference between the
  Atmospheric and Solar Oscillation Amplitudes}},
  \href{https://doi.org/10.1103/PhysRevD.101.093002}{\emph{Phys. Rev. D}
  {\bfseries 101} (2020) 093002}
  [\href{https://arxiv.org/abs/1912.02426}{{\ttfamily 1912.02426}}].

\bibitem{Ioannisian:2020isl}
A.~Ioannisian, S.~Pokorski, J.~Rosiek and M.~Ryczkowski, \emph{{Analytical
  description of CP violation in oscillations of atmospheric neutrinos
  traversing the Earth}},
  \href{https://doi.org/10.1007/JHEP10(2020)120}{\emph{JHEP} {\bfseries 10}
  (2020) 120} [\href{https://arxiv.org/abs/2005.07719}{{\ttfamily
  2005.07719}}].

\bibitem{Minakata:2020ijz}
H.~Minakata, \emph{{Neutrino amplitude decomposition: Toward observing the
  atmospheric - solar wave interference}},
  \href{https://doi.org/10.1140/epjc/s10052-020-08746-6}{\emph{Eur. Phys. J. C}
  {\bfseries 80} (2020) 1207}
  [\href{https://arxiv.org/abs/2006.16594}{{\ttfamily 2006.16594}}].

\bibitem{Minakata:2020oxb}
H.~Minakata, \emph{{Neutrino amplitude decomposition in matter}},
  \href{https://doi.org/10.1103/PhysRevD.103.053004}{\emph{Phys. Rev. D}
  {\bfseries 103} (2021) 053004}
  [\href{https://arxiv.org/abs/2011.08415}{{\ttfamily 2011.08415}}].

\bibitem{Parke:2020wha}
S.J.~Parke, \emph{{Interplay between the factorization of the Jarlskog
  invariant and location of the solar and atmospheric resonances for neutrino
  oscillations in matter}},
  \href{https://doi.org/10.1103/PhysRevD.103.033003}{\emph{Phys. Rev. D}
  {\bfseries 103} (2021) 033003}
  [\href{https://arxiv.org/abs/2012.07186}{{\ttfamily 2012.07186}}].

\bibitem{Minakata:2021dqh}
H.~Minakata, \emph{{Symmetry finder: A method for hunting symmetry in neutrino
  oscillation}}, \href{https://doi.org/10.1103/PhysRevD.104.075024}{\emph{Phys.
  Rev. D} {\bfseries 104} (2021) 075024}
  [\href{https://arxiv.org/abs/2106.11472}{{\ttfamily 2106.11472}}].

\bibitem{Denton:2021vtf}
P.B.~Denton and S.J.~Parke, \emph{{Parameter symmetries of neutrino
  oscillations in vacuum, matter, and approximation schemes}},
  \href{https://doi.org/10.1103/PhysRevD.105.013002}{\emph{Phys. Rev. D}
  {\bfseries 105} (2022) 013002}
  [\href{https://arxiv.org/abs/2106.12436}{{\ttfamily 2106.12436}}].

\bibitem{Minakata:2021goi}
H.~Minakata, \emph{{Symmetry Finder applied to the 1\textendash{}3 mass
  eigenstate exchange symmetry}},
  \href{https://doi.org/10.1140/epjc/s10052-021-09810-5}{\emph{Eur. Phys. J. C}
  {\bfseries 81} (2021) 1021}
  [\href{https://arxiv.org/abs/2107.12086}{{\ttfamily 2107.12086}}].

\bibitem{Minakata:2022yvs}
H.~Minakata, \emph{{Symmetry in Neutrino Oscillation in Matter: New Picture and
  the \ensuremath{\nu}SM\textendash{}Non-Unitarity Interplay}},
  \href{https://doi.org/10.3390/sym14122581}{\emph{Symmetry} {\bfseries 14}
  (2022) 2581} [\href{https://arxiv.org/abs/2210.09453}{{\ttfamily
  2210.09453}}].

\bibitem{Abdullahi:2022fkh}
A.M.~Abdullahi and S.J.~Parke, \emph{{Neutrino Oscillations in Matter using the
  Adjugate of the Hamiltonian}},
  \href{https://arxiv.org/abs/2212.12565}{{\ttfamily 2212.12565}}.

\bibitem{Denton:2023zwa}
P.B.~Denton and J.~Gehrlein, \emph{{Solar parameters in long-baseline
  accelerator neutrino oscillations}},
  \href{https://doi.org/10.1007/JHEP06(2023)090}{\emph{JHEP} {\bfseries 06}
  (2023) 090} [\href{https://arxiv.org/abs/2302.08513}{{\ttfamily
  2302.08513}}].

\bibitem{Minakata:2023ice}
H.~Minakata, \emph{{Neutrino amplitude decomposition, $S$ matrix rephasing
  invariance, and reparametrization symmetry}},
  \href{https://arxiv.org/abs/2308.14501}{{\ttfamily 2308.14501}}.

\bibitem{Huber:2004ka}
P.~Huber, M.~Lindner and W.~Winter, \emph{{Simulation of long-baseline neutrino
  oscillation experiments with GLoBES (General Long Baseline Experiment
  Simulator)}}, \href{https://doi.org/10.1016/j.cpc.2005.01.003}{\emph{Comput.
  Phys. Commun.} {\bfseries 167} (2005) 195}
  [\href{https://arxiv.org/abs/hep-ph/0407333}{{\ttfamily hep-ph/0407333}}].

\bibitem{Huber:2007ji}
P.~Huber, J.~Kopp, M.~Lindner, M.~Rolinec and W.~Winter, \emph{{New features in
  the simulation of neutrino oscillation experiments with GLoBES 3.0: General
  Long Baseline Experiment Simulator}},
  \href{https://doi.org/10.1016/j.cpc.2007.05.004}{\emph{Comput. Phys. Commun.}
  {\bfseries 177} (2007) 432}
  [\href{https://arxiv.org/abs/hep-ph/0701187}{{\ttfamily hep-ph/0701187}}].

\bibitem{Arguelles:2021twb}
C.A.~Arg\"uelles, J.~Salvado and C.N.~Weaver, \emph{{nuSQuIDS: A toolbox for
  neutrino propagation}},
  \href{https://doi.org/10.1016/j.cpc.2022.108346}{\emph{Comput. Phys. Commun.}
  {\bfseries 277} (2022) 108346}
  [\href{https://arxiv.org/abs/2112.13804}{{\ttfamily 2112.13804}}].

\bibitem{Delledalle:2022xx}
C.-A.~Deledalle, L.~Denis and F.~Tupin, \emph{{Speckle Reduction in Matrix-Log
  Domain for Synthetic Aperture Radar Imaging}},
  \href{https://doi.org/10.1007/s10851-022-01067-1}{\emph{J. Math. Imaging
  Vis.} {\bfseries 64} (2022) 298}.

\bibitem{Hahn:2006hr}
T.~Hahn, \emph{{Routines for the diagonalization of complex matrices}},
  \href{https://arxiv.org/abs/physics/0607103}{{\ttfamily physics/0607103}}.

\bibitem{Kopp:2006wp}
J.~Kopp, \emph{{Efficient numerical diagonalization of hermitian 3 x 3
  matrices}}, \href{https://doi.org/10.1142/S0129183108012303}{\emph{Int. J.
  Mod. Phys. C} {\bfseries 19} (2008) 523}
  [\href{https://arxiv.org/abs/physics/0610206}{{\ttfamily physics/0610206}}].

\bibitem{Dziewonski:1981xy}
A.M.~Dziewonski and D.L.~Anderson, \emph{{Preliminary reference earth model}},
  \href{https://doi.org/10.1016/0031-9201(81)90046-7}{\emph{Phys. Earth Planet.
  Interiors} {\bfseries 25} (1981) 297}.

\bibitem{Gonzalo:2023mdh}
T.E.~Gonzalo and M.~Lucente, \emph{{PEANUTS: a software for the automatic
  computation of solar neutrino flux and its propagation within Earth}},
  \href{https://arxiv.org/abs/2303.15527}{{\ttfamily 2303.15527}}.

\bibitem{Coloma:2022umy}
P.~Coloma, M.C.~Gonzalez-Garcia, J.P.~Pinheiro and S.~Urrea,
  \emph{{Constraining new physics with Borexino Phase-II spectral data}},
  \href{https://doi.org/10.1007/JHEP07(2022)138}{\emph{JHEP} {\bfseries 07}
  (2022) 138} [\href{https://arxiv.org/abs/2204.03011}{{\ttfamily
  2204.03011}}]. [Erratum: JHEP 11, 138 (2022)].

\bibitem{Esteban:2020cvm}
I.~Esteban, M.C.~Gonzalez-Garcia, M.~Maltoni, T.~Schwetz and A.~Zhou,
  \emph{{The fate of hints: updated global analysis of three-flavor neutrino
  oscillations}}, \href{https://doi.org/10.1007/JHEP09(2020)178}{\emph{JHEP}
  {\bfseries 09} (2020) 178}
  [\href{https://arxiv.org/abs/2007.14792}{{\ttfamily 2007.14792}}].

\bibitem{nufit-5.2}
I.~Esteban, M.C.~Gonzalez-Garcia, M.~Maltoni, T.~Schwetz and A.~Zhou, ``{NuFIT
  5.2 (2022)}.'' \href{http://www.nu-fit.org}{\tt http://www.nu-fit.org}.

\bibitem{Wolfenstein:1977ue}
L.~Wolfenstein, \emph{{Neutrino Oscillations in Matter}},
  \href{https://doi.org/10.1103/PhysRevD.17.2369}{\emph{Phys. Rev. D}
  {\bfseries 17} (1978) 2369}.

\bibitem{Mikheyev:1985zog}
S.P.~Mikheyev and A.Y.~Smirnov, \emph{{Resonance Amplification of Oscillations
  in Matter and Spectroscopy of Solar Neutrinos}}, {\emph{Sov. J. Nucl. Phys.}
  {\bfseries 42} (1985) 913}.

\bibitem{MINOS:2014rjg}
{\scshape MINOS} collaboration, \emph{{Combined analysis of $\nu_{\mu}$
  disappearance and $\nu_{\mu} \rightarrow \nu_{e}$ appearance in MINOS using
  accelerator and atmospheric neutrinos}},
  \href{https://doi.org/10.1103/PhysRevLett.112.191801}{\emph{Phys. Rev. Lett.}
  {\bfseries 112} (2014) 191801}
  [\href{https://arxiv.org/abs/1403.0867}{{\ttfamily 1403.0867}}].

\bibitem{NOvA:2021nfi}
{\scshape NOvA} collaboration, \emph{{Improved measurement of neutrino
  oscillation parameters by the NOvA experiment}},
  \href{https://doi.org/10.1103/PhysRevD.106.032004}{\emph{Phys. Rev. D}
  {\bfseries 106} (2022) 032004}
  [\href{https://arxiv.org/abs/2108.08219}{{\ttfamily 2108.08219}}].

\bibitem{T2K:2023smv}
{\scshape T2K} collaboration, \emph{{Measurements of neutrino oscillation
  parameters from the T2K experiment using $3.6\times10^{21}$ protons on
  target}},  \href{https://arxiv.org/abs/2303.03222}{{\ttfamily 2303.03222}}.

\bibitem{Wendell:2014dka}
{\scshape Super-Kamiokande} collaboration, \emph{{Atmospheric Results from
  Super-Kamiokande}}, \href{https://doi.org/10.1063/1.4915569}{\emph{AIP Conf.
  Proc.} {\bfseries 1666} (2015) 100001}
  [\href{https://arxiv.org/abs/1412.5234}{{\ttfamily 1412.5234}}]. slides
  available at
  \url{https://indico.fnal.gov/event/8022/other-view?view=standard}.

\bibitem{Kopp:2013vaa}
J.~Kopp, P.A.N.~Machado, M.~Maltoni and T.~Schwetz, \emph{{Sterile Neutrino
  Oscillations: The Global Picture}},
  \href{https://doi.org/10.1007/JHEP05(2013)050}{\emph{JHEP} {\bfseries 05}
  (2013) 050} [\href{https://arxiv.org/abs/1303.3011}{{\ttfamily 1303.3011}}].

\end{thebibliography}\endgroup

\end{document}